\documentclass[aps,prb,twocolumn, superscriptaddress,nobibnotes]{revtex4-1}

\usepackage[english]{babel}
\usepackage{amsmath}
\usepackage{amssymb}
\usepackage{graphicx}
\usepackage{hyperref}

\begin{document}

\title{Two-Particle Self-Consistent method for the multi-orbital Hubbard model}

\author{Karim Zantout}
\email{zantout@itp.uni-frankfurt.de}
\affiliation{Institut f\"ur Theoretische Physik, Goethe-Universit\"at
	Frankfurt, 60438 Frankfurt am Main, Germany}
\author{Steffen Backes}
\affiliation{CPHT, CNRS, Ecole Polytechnique, Institut Polytechnique de Paris, F-91128 Palaiseau, France
	Coll\`ege de France, 11 place Marcelin Berthelot, 75005 Paris, France
	European Theoretical Spectroscopy Facility, 91128 Palaiseau, France, Europe}
\author{Roser Valent\'i}
\affiliation{Institut f\"ur Theoretische Physik, Goethe-Universit\"at Frankfurt, 60438 Frankfurt am Main, Germany}

\begin{abstract}
 	One of the most challenging problems in solid state systems is the
microscopic analysis of electronic correlations. A paramount minimal model that
encodes correlation effects is the Hubbard Hamiltonian, which
-albeit its simplicity- is exactly solvable only in a few limiting cases
and approximate many-body methods are required for its solution.
In this
review we present an overview on the non-perturbative
Two-Particle Self-Consistent method (TPSC) which was 
originally introduced to describe 
the electronic properties of the 
single-band Hubbard model. 
We introduce here a detailed derivation of the multi-orbital
generalization of TPSC and discuss
particular features of the method on exemplary
interacting models in comparison to dynamical mean-field theory results. 
\end{abstract}

\maketitle

\section{Introduction}
In strongly correlated electronic systems the development
of many-body techniques is driven by the fact that 
a description of electronic properties in terms 
of an independent electron picture fails. 
Correlation effects result in a plethora of fascinating phenomena
such as unconventional
superconductivity~\onlinecite{Anderson1987,Kotliar1988,Lee2006,Honerkamp2008,Borisenko2010,WangScience2011,Norman2011,Jerome2012,mazin2014,Steglich2016,Cao2020},
Mott metal-to-insulator
transition~\onlinecite{Mott1982,Imada1998,Powell2011,Raghu2008,bijelic2018,lechermann2018,Roy2019,phillips2010}, non-Fermi
liquid behavior~\onlinecite{Schofield1999,Stewart2001,Lee2018}, or spin liquid
phases~\onlinecite{Balents2010,witczak2014,norman2016,Savary2017,Zhou2017,riedl2019,takagi2019} to mention a few.
In many materials, 
correlations originate from a few partially filled orbitals around the
Fermi level and, early on,
a simplified low-energy description of those orbitals was proposed in terms of the Hubbard 
model~\onlinecite{Gutzwiller1963,Hubbard1963,Kanamori1963,Imada1998}, which maps the electronic part of
the full Hamiltonian of the interacting system
onto an effective lattice model. This model
is expected to capture the correlation effects of the original system but which is still too complex to be solved in the general case except for certain limits.
Thus, one requires the development
of elaborate approximate many-body methods.

Many promising schemes conceived to describe
the electronic properties of correlated materials
start from an \textit{ab-initio}-derived effective non-interacting
Hamiltonian where strong correlation effects are then added and treated
within an approximate many-body method.
Since most recent materials of interest are multi-orbital 
systems~\onlinecite{Salomon2001,singh2009,hirschfeld2011,Fernandes2016,guterding2017,hosono2018,Brahlek2017},
an explicit multi-orbital formulation of many-body techniques is required.

Among the large variety of available many-body methods, in this review  we will focus 
on the so-called conserving approximations in the Baym-Kadanoff sense~\onlinecite{Baym1961,Baym1962}.
Those methods are thermodynamically consistent, 
i.e. thermodynamic expectation values can be obtained as derivatives of the free energy,
and preserve many important physical constraints like the Ward identities for the collective modes
and conservation laws for single-particle properties.
Still, they can differ from one another in how far they 
fulfill other physical constraints like local spin and charge sum rules or the Mermin-Wagner theorem~\onlinecite{VilkTremblay1997}.
In what follows we shortly review
some of these methods.

A very powerful and successful combination of \textit{ab-initio} and many-body techniques is density functional theory 
plus dynamical mean field theory (DFT+DMFT)\onlinecite{Anisimov1997,Lichtenstein1998,held2006} where DFT provides a reasonable starting point for the electronic structure of the system and DMFT~\onlinecite{metzner1989,georges1992,Georges1996} introduces all correlation effects that appear in terms of a dynamical (frequency-dependent) but local self-energy $\Sigma(\omega)$.
Approximating the full momentum- and frequency-dependent self-energy by a local dynamical quantity amounts to 
restricting all correlation effects beyond DFT to a single site.
While this local approximation has been very successful in explaining many experimentally observable
properties of strongly correlated systems~\onlinecite{Georges1996,Kotliar2004,biermann2005,Kotliar2006,yin2011,georges2013,watson2017,backes2015,Vollhardt2019},
there are materials where non-local correlations are non-negligible. This is the case, for instance, close to a phase transition,
or in phenomena like the
pseudo-gap physics in high-Tc cuprates~\onlinecite{Norman1997,Ronning1998,Sadovskii2001,Imada2013,Kordyuk2015}.

A straightforward way of including non-local correlation effects in the DMFT framework are 
the cluster DMFT method (CDMFT)\onlinecite{Georges1996,Kotliar2001,Maier2005,Tremblay2006,Park2008,senechal2015} or the dynamical cluster approximation (DCA)\onlinecite{Hettler1998,Hettler2000,Maier2000,Maier2005,Tremblay2006,lee2012dynamical,Tocchio2013}, which explicitly treat short-range correlations between neighbouring sites by enlarging the unit cell to comprise multiple atoms 
of the same type, but are restricted in spatial resolution due to the large computational cost.

In general, many full momentum-dependent approximations that directly operate in the thermodynamic limit are available, 
both perturbative and non-perturbative in different quantities~\onlinecite{Maier2005,Rohringer2018,Schaefer2020}.
Most straightforward weak-coupling perturbative expansions in the electron-electron interaction
approximate the one-particle irreducible vertex, i.e. the self-energy $\Sigma$.
This one-particle vertex describes the renormalization of an electron due to the electron-electron interaction in the background of 
all other electrons arising from scattering processes.
Such an expansion can also be done in other quantities like the screened interaction $W$. This
is the case in the GW approximation~\onlinecite{Hedin1965,Aryasetiawan1998,Aulbur2000,Reining2018} where the Dyson equation relates the unrenormalized single-particle Green's function $G_0$ and
the single-particle vertex $\Sigma$ with the renormalized Green's function $G$.

Another approach is to expand $\Sigma$ non-perturbatively in the interaction,
but perturbatively in the locality of the diagrams~\onlinecite{Rohringer2018}.
The DMFT approximation is then the lowest order term in the sense of locality, 
since it approximates the one-particle vertex $\Sigma$ to be local, but generates it from
a summation of all diagrams that can be obtained from local propagators.

An alternative procedure is
to approximate two- particle quantities like the 
irreducible vertex $\Gamma$, from which one-particle quantities can be derived that usually contain a richer structure of correlation effects.
On the two-particle level, the irreducible vertex $\Gamma$ contains information about
two-particle scattering processes. Here, the Bethe-Salpeter equation represents a 
two-particle analogue of the single-particle Dyson equation, relating two-particle Green's functions like the bare  and renormalized generalized susceptibilities with the two-particle irreducible vertex~\onlinecite{Salpeter1951,Kadanoff1962}.
Methods like the Random Phase Approximation (RPA) or the Fluctuation Exchange Approximation (FLEX)~\onlinecite{Bickers1989,BickersPRL1989,BickersPRB1991,Bickers1991,graser2009,altmeyer2016role}
sum certain subsets of diagrams to approximate the two-particle vertex, 
while D$\Gamma$A~\onlinecite{Toschi2007,Held2008,Rohringer2018} approximates the vertex as a dynamical but local quantity, including all local diagrams.
Further two-particle extensions for the vertex are for example TRILEX~\onlinecite{Ayral2015,Ayral2016}, QUADRILEX~\onlinecite{AyralQUADRILEX2016}, dual boson~\onlinecite{Rubtsov2012} and dual fermion techniques~\onlinecite{Rubtsov2009} or GW+DMFT~\onlinecite{Biermann2003,Biermann2014}
which use the local DMFT solution and vertex as a starting point for a perturbation series to generate non-local diagrams.

In this review we focus on the Two-Particle Self-Consistent approach (TPSC).
This is a method developed within the Baym-Kadanoff scheme~\onlinecite{Baym1961,Kadanoff1962,Baym1962,Bickers2004,Tremblay2019} that retains the dynamical and non-local nature of electronic correlations, 
while using physical sum rules to obtain consistent values for all the quantities that are approximated.
As described in Section~\ref{sec:single}, one approximates the two-particle irreducible vertex $\Gamma$, usually depending
on three frequency- and momentum-indices~\onlinecite{Rohringer2012,Schaefer2013,Hafermann2014,Janis2014,Gunnarsson2016,Schaefer2016,Schaefer2020}, to be frequency and momentum independent, i.e. the two-particle irreducible vertex $\Gamma$ is a mere constant.
The vertex $\Gamma$ is then determined by requiring the spin and charge susceptibilities to obey physical summation rules. This is in contrast to many
approaches where the (approximate) irreducible vertices  are obtained by solving complicated Parquet equations~\onlinecite{Rohringer2018,Schaefer2020}. 
From the equations of motion derived for the Green's function (also called Schwinger-Dyson equation) the self-energy can then be obtained from the bare interaction, bare Green's function, two-particle irreducible vertices and generalized susceptibilities.
This local and static approximation of the vertex $\Gamma$ in the two-particle sector
results in a non-perturbative, fully frequency and momentum dependent single-particle 
self-energy $\Sigma$, which has been shown to be able to describe many electronic
correlation effects including pseudo-gap physics and superconductivity\onlinecite{Tremblay2006,Tremblay2012}
as will be discussed in section~\ref{sec:history}.

In this review we present a detailed derivation of an extension of the standard single-orbital Two-Particle Self-Consistent method to the multi-orbital case as presented in~\onlinecite{Zantout2019} and discuss applications to model systems, as well as benchmarks with other methods.

\section{Single-band TPSC method \label{sec:single}}
\begin{figure}
	\centering
	\includegraphics[width=0.45\textwidth]{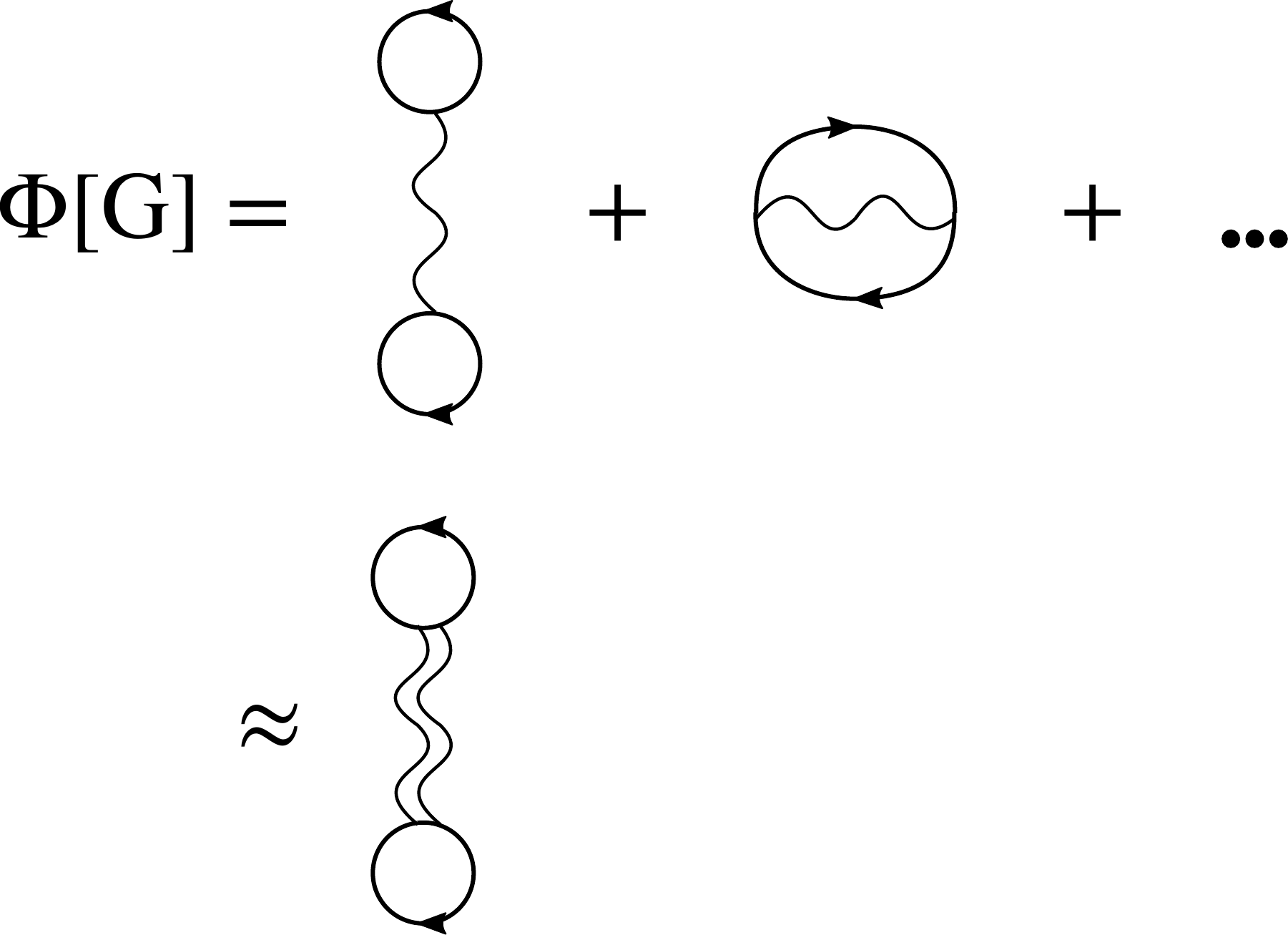}%
	\caption{\label{fig: diag}
		The Luttinger-Ward functional $\Phi[G]$ is by definition the sum of all closed two-particle irreducible skeleton diagrams. In the diagrammatic representation the bold lines correspond to full Green's functions while the single-wiggled lines are interaction vertices. In the TPSC approximation one assumes that all diagrams can be approximated by a diagram of first order where the interaction vertex is replaced by an effective irreducible interaction vertex (double-wiggled line).}
\end{figure}

Before we outline the history of the TPSC method and present a detailed derivation of the multi-orbital TPSC we would like to sketch in this section the main ideas of the single-band TPSC~\onlinecite{VilkTremblay1997,Tremblay2012} that is formulated for a Hubbard model with on-site interaction $U$.\\
Since TPSC is derived within the Baym-Kadanoff scheme~\onlinecite{Baym1961,Baym1962,Kadanoff1962,Bickers2004,Tremblay2019}, we start with the description of the
interacting system in terms of a Luttinger-Ward functional
$\Phi[G]$~\onlinecite{Luttinger1960,LuttingerWard1960,Potthoff2003},
which is a scalar functional of the dressed many-body Green's function $G$. Specifically,
$\Phi[G]$ incorporates all closed two-particle irreducible skeleton diagrams
constructed from $G$ and the on-site interaction $U$. In
general, however, $\Phi[G]$ cannot be evaluated explicitly but one can
approximate it which is the idea of conserving approximations. In TPSC one
assumes that at intermediate interaction strengths one can absorb the effect of
diagrams of all orders into an effective irreducible interaction four-point
vertex $\Gamma$ that is local and static and appears only in the first order
diagram (see fig.~\ref{fig: diag}).

The effective irreducible interaction vertex $\Gamma$ can be decomposed into a spin vertex $\Gamma^{sp}$ and a charge vertex $\Gamma^{ch}$. These two vertices are then determined from the spin and charge susceptibilities $\chi^{sp/ch}$; more specifically from the so-called local spin and charge sum rules,
\begin{align}
\chi^{sp}(\vec r=0,t =0) &= \langle n\rangle -2\langle n_\uparrow n_\downarrow\rangle\label{eq: sp_sumrule_1band}\\
\chi^{ch}(\vec r=0,t =0) &= \langle n\rangle +2\langle n_\uparrow n_\downarrow\rangle -\langle n\rangle^2
\end{align}
where $\langle n\rangle $ is the filling and $\langle n_\uparrow n_\downarrow\rangle$ is the double occupation. The single-band TPSC approach needs an ansatz for the calculation of the spin vertex $\Gamma^{sp}$ or, equivalently, of the double occupancy $\langle n_\uparrow n_\downarrow\rangle$ given by
\begin{equation}
\Gamma^{sp} = \frac{\langle n_\uparrow n_\downarrow\rangle}{n_\uparrow n_\downarrow} U.\label{eq: Usp_ansatz_1band}
\end{equation}
Having determined the spin and charge vertices $\Gamma^{sp},~\Gamma^{ch}$ one uses the Bethe-Salpeter equation~\onlinecite{Salpeter1951,Kadanoff1962} for the two-particle Green's function to compute
\begin{align}
\chi^{sp} &= 2\chi^0/\left(1-\chi^0\Gamma^{sp}\right)\label{eq: chi_sp_1band}\\
\chi^{ch} &= 2\chi^0/\left(1+\chi^0\Gamma^{ch}\right)
\end{align}
where $\chi^0$ is the particle-hole bubble diagram $-G^0\ast G^0$ calculated from the non-interacting Green's function $G^0$.\\
By construction, the self-energy $\Sigma$ is computed from $\Sigma=\frac{\delta
	\Phi[G]}{\delta G}$ which equals some constant in the case of the TPSC
Luttinger-Ward functional. This constant can be absorbed into the chemical
potential and therefore no single-particle renormalizations take place. In the framework of conserving approximations one can further improve on this result by using the Bethe-Salpeter equation for the
self-energy~\onlinecite{Kadanoff1962,VilkTremblay1997},
\begin{align}
\Sigma = \Sigma_{HF} + \frac{U}{4}\left[\chi^{sp}\Gamma^{sp}+\chi^{ch}\Gamma^{ch}\right]\ast G^{0},
\end{align}
where the non-interaction Green's function $G^0$ appears instead of the dressed Green's function $G$ to preserve consistency with the TPSC Luttinger-Ward functional and
where $\Sigma_{HF}$ is the Hartree-Fock self-energy and $\ast$ denotes the convolution. This improvement yields an approximation that is not conserving in the strict Kadanoff-Baym sense~\onlinecite{Baym1961,Kadanoff1962,Baym1962} but still retains conservation laws to a high degree~\onlinecite{VilkTremblay1997}.\\
Additionally, one can further improve the self-energy by taking crossing symmetry of the two-particle irreducible vertex $\Gamma$ into account~\onlinecite{Moukouri2000} which gives
\begin{equation}
\Sigma = \Sigma_{HF} +
\frac{U}{8}\left[3\chi^{sp}\Gamma^{sp}+\chi^{ch}\Gamma^{ch}\right]\ast G^{0}.
\end{equation}
Finally, one uses the Dyson equation
\begin{equation}
G^{-1} = (G^0)^{-1}-\Sigma
\end{equation}
to obtain the full Green's function $G$.

\section{\label{sec:history}Previous formulations of the TPSC method}
In this section we provide a brief overview on past developments of TPSC.\\
Early on, prior to the formulation of the two-particle self-consistent method, Vilk \textit{et al.} introduced in~\onlinecite{Vilk1994} a simple self-consistent way of obtaining approximate spin and charge susceptibilities of the two-dimensional one-band Hubbard model without adjustable parameters. The ansatz was motivated by the local field approximation~\onlinecite{Singwi1968,Ichimaru1982}. 
This approach provided results comparable to quantum Monte Carlo simulations at weak to intermediate coupling strength and respected the Mermin-Wagner theorem that prohibits in two dimensions a spontaneous breaking of the $SO(3)$ spin symmetry at finite temperature. 
A few advantages of this ansatz, which are also present in the TPSC equations, are the inclusion of short-range quantum fluctuations and finite temperature effects and the ability to reproduce Kanamori-Brueckner (KB) screening, which 
describes the reduction of the bare interaction in the spin channel
~\onlinecite{Kanamori1963}. 
First applications to the single-band Hubbard model far from van Hove singularities revealed valuable insights into the spin and charge fluctuations~\onlinecite{Vilk1994,VilkChen1994,Veilleux1995,Nelisse1999}.



The first complete formulation of single-band TPSC was introduced in~\onlinecite{Vilk1996} and a very extensive and thorough presentation of it can be found in \onlinecite{VilkTremblay1997,Allen2001,Tremblay2012}.

The TPSC method was developed with the aim of fulfilling essential physical properties such as the local spin and charge sum rules, 
which is, for instance, violated by the FLEX approximation, 
and conservation laws of the spin and charge susceptibilities, like $\chi_{sp/ch}(\mathbf{q}=0,iq_n)=0$ for $q_n\neq 0$,
which are also not fulfilled in the FLEX approximation.
Actually, TPSC not only fulfills the Mermin-Wagner theorem and local spin and charge sum rules exactly, but it also satisfies Luttinger's theorem and the f-sum rule to a high degree (the deviation is of the order of a few percent)~\onlinecite{VilkTremblay1997}.
Extensions of TPSC including the transversal particle-hole channel
contributions to the electronic self-energy~\onlinecite{Moukouri2000,Allen2001} preserve also the crossing symmetry of the irreducible four-point vertex $\Gamma$, while obeying spin rotational invariance.
The single-band TPSC method has been successfully applied to a multitude of
physical phenomena described by the single-band Hubbard 
model~\onlinecite{Vilk1995,Vilk1996,Dare1996,Vilk1997,VilkTremblay1997,Dare2000,Bergeron2012,Dare2007}.
For instance, mostly in the framework of high-Tc cuprate superconductors,
studies on the precursor antiferromagnetic bands, the pseudogap
phase, superconducting transition temperatures, spectral and dynamical properties~\onlinecite{Kyung2003,KyungPRL2003,Hankevych2003,Kyung2004,Hassan2008,Hankevych2006,Ogura2015,Nishiguchi2017,Nishiguchi2018} and
extent of quantum criticality~\onlinecite{Roy2008,Bergeron2011} provided further insight on these materials and model systems.
Specifically, this approach was used to study universal critical behavior, where TPSC was shown to be in the same universality class as the $n\rightarrow\infty$ limit of the $O(n)$ classical vector model~\onlinecite{Dare1996} while $n=3$ would correspond to the correct limit for a magnetic transition.
The proof in the referenced publication is cutoff independent and has a lower critical dimension $d=2$, where the correlation length grows exponentially. 
TPSC was used to estimate the extent of quantum criticality in the Hubbard model phase diagram~\onlinecite{Roy2008} with applications to high-Tc cuprate superconductors~\onlinecite{Bergeron2012}.
Interestingly, the same critical behavior is also observed~\onlinecite{DelRe2019} in a special formulation of the D$\Gamma$A which might be due to a possible similarity of both methods, where TPSC (static vertex function) is some limiting case of D$\Gamma$A (dynamical vertex function).

However, TPSC fails to be a good approximation in the strong-coupling regime
where the frequency and momentum dependence of the irreducible four-point
vertex $\Gamma$ becomes important while the TPSC vertex is, by construction,
constant.  The method is therefore not able to describe Hubbard
satellites, but only precursors of antiferromagnetic bands and
pseudogaps~\onlinecite{VilkTremblay1997}.
Only very recently, the so-called TPSC+ approach was introduced in~\onlinecite{Schaefer2020} where effective frequency-dependent vertex corrections were included in the TPSC equations to improve the results in the intermediate coupling regime.

The efficiency and reliability of TPSC has been discussed in comparison to a few other
approximations like paramagnon and perturbation theories~\onlinecite{Saikawa2001}, the dynamical cluster approximation~\onlinecite{KyungPRL2003} and the FLEX
approximation~\onlinecite{Arita2000}.
For more detailed comparisons of TPSC to other many-body method we point to~\onlinecite{Schaefer2020,Rohringer2018}.
In this context, in Ref.~\onlinecite{Mertz2018} a measure of the self-energy dispersion was defined emphasizing the role of local versus non-local correlations taking
as reference TPSC versus 
DMFT calculations. Based on this concept, non-local correlation effects on topological properties in the Haldane-Hubbard model~\onlinecite{Mertz2019}
and the bilayer ''twistronic`` 1T-TaSe$_2$~\onlinecite{Pizarro2020} have recently been investigated.

The single-band TPSC has been also applied to the attractive Hubbard model, where spin and charge fluctuations are replaced by pairing fluctuations. The general scheme of the derivation remains identical, and the conserving and respective single- and two-particle properties are retained~\onlinecite{Allen2000,Kyung2000,AllenPRB2001,Verga2005,Otsuki2012}. The testing ground for this TPSC implementation was found in underdoped cuprates, with a focus on the interplay between superconductivity and pseudogap physics~\onlinecite{Kyung2000,Kyung2001} and good agreement to QMC data in the weak to intermediate coupling regime was reported~\onlinecite{Kyung2001}.

Further extension of TPSC included the effect of nearest-neighbour interactions $V$ in the extended single-band Hubbard model~\onlinecite{Davoudi2006}
and the effect of pair- correlation functions derivatives which had been neglected so far~\onlinecite{Davoudi2007,Davoudi2008}.This formulation yields not only good agreement with QMC when $V$ is small, but also in the limit where charge fluctuations are the main contribution, in contrast to FLEX and mean-field calculations.

Extensions to multi-site TPSC were introduced in Ref.~\onlinecite{Arya2015} and applied to study
the semimetal to antiferromagnetic phase transition
on a honeycomb lattice~\onlinecite{Arya2015} and,
the metal to superconducting transition in organic superconductors~\onlinecite{Zantout2018,Aizawa2015,Aizawa2018}.

The first extension of TPSC to a multi-orbital formulation was introduced by Miyahara \textit{et al.}~\onlinecite{Miyahara2013}
where superconducting critical temperatures and gap symmetries with additional comparisons to RPA and FLEX were performed for high-Tc superconductors.

The authors of this review introduced in~\onlinecite{Zantout2019} an alternative multi-orbital formulation of TPSC that is presented in the next section.
We will discuss the differences between both multi-orbital formulations in the next section.
An extended study~\onlinecite{Bhattacharyya2020} on a larger class of iron-based superconductors 
focussed on a comparison of multi-orbital TPSC with FLEX and RPA and underlined the importance of non-local effects in those materials.


\section{Derivation of multi-orbital TPSC}

In this section we present a detailed derivation of multi-orbital TPSC~\onlinecite{Zantout2019}. The aim is to formulate an approximation to solve the multi-orbital Hubbard Hamiltonian that does not violate conservation laws. 
To do so one starts from the Luttinger-Ward functional~\onlinecite{LuttingerWard1960,Luttinger1960} and applies approximations on the four-point vertex as in fig.~\ref{fig: diag}.
In a conserving approximation one restricts all possible closed skeleton two-particle diagrams, i.e. diagrams that contain fully dressed Green's functions without explicit self-energy lines, to a subset of diagrams.
From functional differentiation it is then possible to compute a self-energy that is consistent with the chosen set of skeleton diagrams.
If the chosen subset of diagrams leads to convergent series with a physical solution, 
one can be sure to have a conserving approximation in the Baym-Kadanoff sense.
All expressions are given in Planck units where $\hbar = k_B = 1$.
\subsection{Definitions}
The full Hamiltonian that is considered in this work is given by 
\begin{align}
H =& \sum_{\alpha,\beta, i, j,\sigma}\left(t_{\alpha\beta}^{\vec R_i-\vec R_j}-\mu_0\delta^{}_{i,j}\delta^{}_{\alpha,\beta}\right)c_{\alpha,\sigma}^\dag(\vec R_i)c^{}_{\beta,\sigma}(\vec R_j) \nonumber\\
+& \frac{1}{2}\sum_{\alpha,\beta,i,\sigma}U_{\alpha\beta}n_{\alpha,\sigma}(\vec R_i)n_{\beta,-\sigma}(\vec R_i)\nonumber\\
+& \frac{1}{2}\sum_{\substack{\alpha,\beta,i,\sigma \\ \alpha \neq \beta}}(U_{\alpha\beta}-J_{\alpha\beta})n_{\alpha,\sigma}(\vec R_i)n_{\beta,\sigma}(\vec R_i) \nonumber\\
-&\frac{1}{2}\sum_{\substack{\alpha,\beta,i,\sigma \\ \alpha \neq \beta}}J_{\alpha\beta}\left(c_{\alpha,\sigma}^\dag(\vec R_i)c^{}_{\alpha,-\sigma}(\vec R_i)c^{\dag}_{\beta,-\sigma}(\vec R_i)c^{}_{\beta,\sigma}(\vec R_i)\right.\nonumber\\
+&\left.c^{\dag}_{\alpha,\sigma}(\vec R_i)c^{}_{\beta,-\sigma}(\vec R_i)c^{\dag}_{\alpha,-\sigma}(\vec R_i)c^{}_{\beta,\sigma}(\vec R_i)\right),\label{Eq-Hubbard-H}
\end{align}
where $t_{\alpha\beta}^{\vec R_i-\vec Rj}$ are all hoppings concerning orbitals $\alpha$ and $\beta$ that are connected by lattice vectors $\vec R_i - \vec R_j$  and we dropped the spin index since we assume a paramagnetic state without breaking of time-reversal symmetry. $U_{\alpha\beta}$ denote the onsite orbital-dependent Hubbard interactions and $J_{\alpha\beta}$ are the onsite inter-orbital Hund's couplings. 
The operator $c^{}_{\alpha,\sigma}(\vec R_i,\tau)$ destroys an electron with spin $\sigma$ in the $\alpha$-orbital at unit cell position $\vec R_i$ at the imaginary time $\tau$ and $c^\dag_{\beta,\sigma}(\vec R_j,\tau')$ creates an electron with spin $\sigma$ in the $\beta$-orbital at unit cell position $\vec R_j$ at $\tau'$ and $n^{}_{\alpha,\sigma}(\vec R_i,\tau):=c^{\dag}_{\alpha,\sigma}(\vec R_i,\tau)c^{}_{\alpha,\sigma}(\vec R_i,\tau)$. Note that we dropped the time dependence of the Hamilton operator since it is not explicitly time dependent.\\
The multi-orbital Green's function for a lattice system is defined as
\begin{align}
G_{\mu\nu,\sigma}(\vec R_i,\tau;\vec R_j,\tau') := -\langle T_\tau c^{}_{\mu,\sigma}(\vec R_i,\tau)c^\dag_{\nu,\sigma}(\vec R_j,\tau') \rangle.
\end{align}

Due to space-time translational invariance one can rewrite
\begin{align*}
G_{\mu\nu,\sigma}(\vec R_i,\tau;\vec R_j,\tau') &= G_{\mu\nu,\sigma}(\vec R_i-\vec R_j,\tau-\tau';0,0)\\
&=: G_{\mu\nu,\sigma}(\vec R_i-\vec R_j,\tau-\tau').
\end{align*}
Taking advantage of the spatial translational invariance the Fourier transform of the expression above yields the Green's function in reciprocal space $G_{\mu\nu,\sigma}(\vec k,\tau-\tau')$, where $\vec k$ is a reciprocal lattice vector.
Furthermore, one uses the anti-periodicity of the Green's function $G_{\mu\nu,\sigma}(\vec k,\tau)$ in imaginary time $\tau$ with period $\tilde{\beta}:=\frac{1}{k_BT}$,
\begin{align}
G_{\mu\nu,\sigma}(\vec k,\tau) &= -G_{\mu\nu,\sigma}(\vec k,\tau-\tilde\beta)\hspace{1cm}\forall \tau \in(0,\tilde\beta),\label{Eq-G-antiperiod}
\end{align}
to define the Matsubara Green's function
\begin{align}
G_{\mu\nu,\sigma}(\vec k,i\omega_n) &:= \int_0^{\tilde\beta} e^{i\omega_n \tau}G_{\mu\nu,\sigma}(\vec k,\tau)d\tau\nonumber\\
G_{\mu\nu,\sigma}(\vec k,\tau) &= \frac{1}{\tilde\beta}\sum_{n=-\infty}^{\infty}e^{-i\omega_n \tau}G_{\mu\nu,\sigma}(\vec k,i\omega_n),
\end{align}
where the fermionic Matsubara frequencies are defined by  $\omega_n := \frac{(2n+1)\pi}{\tilde\beta}$.

The kinetic part of the Hamilton operator
\begin{align}
H^{kin}_{\alpha\beta}(\vec R_i,\vec R_j) &= \left(t_{\alpha\beta}^{\vec R_i-\vec R_j}-\mu_0\delta_{\vec R_i,\vec R_j}\delta_{\alpha,\beta}\right)c_{\alpha}^\dag(\vec R_i)c^{}_{\beta}(\vec R_j) \nonumber\\
&= \sum_{\vec k,\vec k'} \left(t_{\alpha\beta}^{\vec R_i-\vec R_j}-\mu_0\delta_{\vec R_i,\vec R_j}\delta_{\alpha,\beta}\right)\cdot\nonumber\\
&\hspace{0.8cm}\cdot e^{-i\vec k\vec R_i}c_{\alpha}^\dag(\vec k)e^{i\vec k'\vec R_j}c^{}_{\beta}(\vec k')
\end{align}
can be diagonalized via a Fourier transformation which leads to a set of eigenvectors $a_b(\vec k)$ and corresponding eigenvalues $\xi_b(\vec k)$, where $b$ is the numeration index (band index), i.e.
\begin{align}
h^{kin}(\vec k) &= \left[a_1(\vec k)\cdots a_{N_{\text{orb}}}(\vec k)\right] \begin{bmatrix}
\xi_1(\vec k) & & \\
& \ddots  &  \\
& & \xi_{N_\text{orb}}(\vec k)
\end{bmatrix} \cdot\nonumber\\
&\hspace{0.2cm}\cdot\left[a_1(\vec k) \cdots a_{N_\text{orb}}(\vec k)\right]^\dag,\label{Eq: Hamiltonian_trafo}
\end{align}
where $N_\text{orb}$ is the number of orbitals and 
\begin{equation}
H_{\alpha\beta}^{kin}(\vec k)=c_{\alpha}^\dag(\vec k)h^{kin}_{\alpha\beta}(\vec k)c^{}_{\beta}(\vec k).
\end{equation}
Thus, the non-interacting Matsubara Green's function reads 
\begin{align}
\left(G^0\right)_{\mu\nu}(\vec k,i\omega_n) &= \left[i\omega_n \mathbf{I} - h^{kin}(\vec k)\right]^{-1}_{\mu\nu}.
\end{align}
From the Matsubara Green's function we can calculate the filling $\langle n\rangle$ via
\begin{align}
\langle n\rangle &= \frac{1}{\tilde\beta N_\text{orb}N_{\vec k}}\lim_{\tau\rightarrow0^-}\sum_\mu\sum_{\vec k,\sigma} \sum_{n=-\infty}^{\infty}e^{-i\omega_n\tau}G_{\mu\mu,\sigma}(\vec k,i\omega_n).\label{Eq-filling}
\end{align}

To abbreviate the notation we introduce the following convention for space-time vectors and functions:
\begin{align}
(\vec R_1,\tau_1)&:= (1)\nonumber\\
f(\vec R_1,\tau_1)&:= f(1)\nonumber\\
f(\vec R_1-\vec R_2,\tau_1-\tau_2)&:= f(1-2)\label{Eq-abbrev}
\end{align}
The multi-orbital non-interacting susceptibility is defined as
\begin{align}
\chi^0_{\lambda\mu\nu\xi}(1-2) &:= -G_{\mu\xi}^0(1-2)G_{\nu\lambda}^0(2-1)\nonumber\\
\chi^0_{\lambda\mu\nu\xi}(\vec q,iq_m) &= -\frac{1}{\tilde\beta N_{\vec k}}\sum\limits_{\vec k,i\omega_n}G_{\nu\lambda}^0(\vec k,i\omega_n)G_{\mu\xi}^0(\vec k+\vec q,i\omega_{n+m}),\label{Eq: chi0}
\end{align}
where the bosonic Matsubara frequencies are given by $q_n :=2\pi n/\tilde\beta$.
The non-interacting susceptibilities are 4-index-tensors. 
We have dropped the spin index since the non-interacting Hamiltonian is spin-rotational invariant.

\subsection{Functional Derivative Approach for the longitudinal channel in the multi-orbital Hubbard model}
In order to be able to include the orbital-dependent interactions $U_{\mu\nu},~J_{\mu\nu}$ in the preceding formalism one introduces an artificial scalar field $\phi_{\mu\nu,\sigma}(1)$ that depends on the orbitals $\mu,\nu$ and the position and imaginary time coordinate $1$ (see eqs.~\eqref{Eq-abbrev}) as developed by Kadanoff and Baym in~\onlinecite{Baym1961,Baym1962}.\\
This field is supposed to couple to the field operators of equal spin (longitudinal channel) like in the generalized partition function in the grand canonical ensemble
\begin{align}
Z[\phi] &= tr\left[e^{-\tilde\beta H}T_\tau e^{-\sum\limits_{\mu\nu,\sigma}\int d(1) c^\dag_{\mu,\sigma}(1)\phi^{}_{\mu\nu,\sigma}(1)c^{}_{\nu,\sigma}(1)}\right] \nonumber\\
&=: \left\langle T_\tau e^{-c^\dag_{\bar\mu,\bar\sigma}(\bar 1)\phi^{}_{\bar{\mu}\bar\nu,\bar\sigma}(\bar 1)c^{}_{\bar\nu,\bar\sigma}(\bar 1)}\right\rangle \nonumber\\
&=: \left\langle T_\tau S[\phi]\right\rangle,\label{Eq-Z}
\end{align} 
where we use a short-hand notation for integrations and summation,
i.e. symbols with a bar are summed or integrated over.
Such a coupling that leads to spin and charge fluctuations in the particle-hole channel only are the most important modes in the repulsive Hubbard model~\onlinecite{Moukouri2000,Allen2001}. 
Although an inclusion of the transversal channel ($\phi$ couples to opposite spins) leads also to quantum fluctuations, it was not considered so far~\onlinecite{Miyahara2013,Zantout2019} (see section~\ref{sec: Outlook}). 

Obviously, one can recover the usual physics by setting the field $\phi$ to zero. The advantages of such a field become visible in the following steps.\\
First, a generalized Green's function $G_\sigma(1,2;[\phi])=:G_\sigma(1,2)_\phi$ can be defined as
\begin{align}
G^{}_{\mu\lambda,\sigma}(1,2)_\phi	&:= -\frac{\delta \ln Z[\phi]}{\delta \phi_{\lambda\mu,\sigma}(2,1)}\nonumber\\
&=- \frac{\left\langle T^{}_\tau S[\phi] c^{}_{\mu,\sigma}(1)c^\dag_{\lambda,\sigma}(2)\right\rangle}{Z[\phi]}\nonumber\\
&=:- \left\langle c^{}_{\mu,\sigma}(1)c^\dag_{\lambda,\sigma}(2)\right\rangle_\phi.\label{Eq-G_defn}
\end{align}
In the framework of conserving approximation also higher-order derivative with respect to the field $\phi$,
\begin{align}
&\left.\frac{\delta G^{}_{\mu\lambda,\sigma}(1,2)_\phi}{\delta\phi^{}_{\xi\nu,\sigma'}(3,4)}\right|_{\phi=0} =\nonumber\\
=& \left\langle c^{}_{\mu,\sigma}(1)c^\dag_{\lambda,\sigma}(2)c^\dag_{\xi,\sigma'}(3)c^{}_{\nu,\sigma'}(4)\right\rangle + G^{}_{\mu\lambda,\sigma}(1,2) G^{}_{\nu\xi,\sigma'}(4,3),\label{Eq-dGdphi}
\end{align}
are considered.
This expression will be an important ingredient in the formulation of a self-consistency.

\subsubsection{The self-energy and Dyson equation}
\label{Sec: Dyson equation}
In this framework the self-energy $\Sigma$ can be defined implicitly via
\begin{align}
&\Sigma_{\alpha\bar\beta,\sigma}(1,\bar 3)_\phi G_{\bar \beta \gamma,\sigma}(\bar 3, 2)_\phi =\nonumber\\
&-\sum_{\beta}U_{\beta\alpha} \left\langle n^{}_{\beta,-\sigma}(1^+)c^{}_{\alpha,\sigma}(1)c^\dag_{\gamma,\sigma}(2)\right\rangle_\phi\nonumber\\
&-\sum_{\substack{\beta \\ \beta\neq\alpha}}(U_{\beta\alpha}-J_{\beta\alpha}) \left\langle n^{}_{\beta,\sigma}(1^+)c^{}_{\alpha,\sigma}(1)c^\dag_{\gamma,\sigma}(2)\right\rangle_\phi\nonumber\\
&+\sum_{\substack{\beta\\\beta\neq\alpha}}J_{\beta\alpha}\left(\left\langle c^\dag_{\beta,-\sigma}(1^+)c^{}_{\beta,\sigma}(1)c^{}_{\alpha,-\sigma}(1)c^\dag_{\gamma,\sigma}(2)\right\rangle_\phi\right.\nonumber\\
&+\left.\left\langle c^\dagger_{\alpha,-\sigma}(1^{++})c^{}_{\beta,\sigma}(1^+)c^{}_{\beta,-\sigma}(1)c^\dag_{\gamma,\sigma}(2)\right\rangle_\phi\right)\label{Eq-self-energy}
\end{align}
and leads to the Dyson equation
\begin{align}
\left(G^{-1}\right)_{\nu\xi,\sigma}(1,2)_\phi =& \left[\left(G^0\right)^{-1}\right]_{\nu\xi}(1,2) \nonumber\\
&- \phi^{}_{\nu\xi,\sigma}(1,2) - \Sigma^{}_{\nu\xi,\sigma}(1,2)_\phi.\label{Eq-Dyson}
\end{align}

\subsubsection{A self-consistent formulation for $\frac{\delta G}{\delta \phi}$}

So far, we only have an implicit equation for the self-energy (eq.~\eqref{Eq-self-energy}) but the right-hand side of the equation is already known from eq.~\eqref{Eq-dGdphi} and, therefore, one can rewrite
\begin{align}
&\Sigma_{\alpha\bar \beta,\sigma}(1,\bar 3) G_{\bar \beta\gamma,\sigma}(\bar 3, 2)= \nonumber\\
& -\sum_{\beta}U_{\beta\alpha}\cdot\nonumber\\
&\hspace{0.6cm}\cdot\left(\left.\frac{\delta G_{\alpha\gamma,\sigma}(1,2)_\phi}{\delta\phi_{\beta\beta,-\sigma}(1^{++},1^+)}\right|_{\phi=0}-G_{\beta\beta,-\sigma}(1,1^{+}) G_{\alpha\gamma,\sigma}(1,2)\right)\nonumber\\
&-\sum_{\substack{\beta \\ \beta\neq\alpha}}(U_{\beta\alpha}-J_{\beta\alpha})\cdot\nonumber\\
&\hspace{0.6cm}\cdot\left(\left.\frac{\delta G_{\alpha\gamma,\sigma}(1,2)_\phi}{\delta\phi_{\beta\beta,\sigma}(1^{++},1^+)}\right|_{\phi=0}-G_{\beta\beta,\sigma}(1,1^{+}) G_{\alpha\gamma,\sigma}(1,2)\right)\nonumber\\
&-\sum_{\substack{\beta\\\beta\neq\alpha}}J_{\beta\alpha}\cdot\nonumber\\
&\hspace{0.6cm}\cdot\left[\left(\left.\frac{\delta G_{\beta\gamma,\sigma}(1,2)_\phi}{\delta\phi_{\beta\alpha,-\sigma}(1^{+},1^{++})}\right|_{\phi=0}-G_{\beta\gamma,\sigma}(1,2) G_{\alpha\beta,-\sigma}(1^{+},1)\right)\right.\nonumber\\
&\left.\hspace{0.6cm}+\left(\left.\frac{\delta G_{\beta\gamma,\sigma}(1^+,2)_\phi}{\delta\phi_{\alpha\beta,-\sigma}(1^{++},1)}\right|_{\phi=0}-G_{\beta\gamma,\sigma}(1,2) G_{\beta\alpha,-\sigma}(1,1^{+})\right)\right]\label{Eq-self-energy-dGdphi},
\end{align}
where the notation $1^{+}$ and $1^{++}$ is used to make the expression well-defined: The ordering of the operators within the functions is defined via $\tau_1^{++}>\tau_1^{+}>\tau_1$ with infinitesimal small differences between them.

The difficult part now is to evaluate the variational differentiation on the right-hand side of equation~\eqref{Eq-self-energy-dGdphi} which can be further simplified with the Bethe-Salpeter equation
\begin{align}
&\frac{\delta G^{}_{\mu\lambda,\sigma}(1,6)_\phi}{\delta\phi^{}_{\xi\nu,\sigma'}(4,5)}\nonumber\\
=& G^{}_{\mu\xi,\sigma}(1,4)_\phi G^{}_{\nu\lambda,\sigma}(5,6)_\phi\delta_{\sigma,\sigma'}\nonumber\\
&+ G^{}_{\mu\bar\beta,\sigma}(1,\bar 3)_\phi\frac{\delta \Sigma^{}_{\bar \beta \bar\gamma,\sigma}(\bar 3,\bar 2)_\phi}{\delta G^{}_{\bar\rho\bar \lambda,\overline{\sigma''}}(\bar 7,\bar 8)_\phi}\frac{\delta G^{}_{\bar\rho\bar \lambda,\overline{\sigma''}}(\bar 7,\bar 8)_\phi}{\delta\phi^{}_{\xi\nu,\sigma'}(4,5)} G^{}_{\bar\gamma\lambda,\sigma}(\bar 2,6)_\phi. \label{Eq-dGdphi-sc}
\end{align}
Before this step we present relationships between $\delta G/\delta \phi$ and spin and charge susceptibilities.

\subsection{Spin and charge susceptibilities}
One defines the charge susceptibility as the linear response to charge perturbations
\begin{align}
\chi^{ch}_{\lambda\mu\nu\xi}(1,2) :=& \langle T_\tau n_{\mu\lambda}(1)n_{\nu\xi}(2)\rangle - \langle n_{\mu\lambda}(1)\rangle\langle n_{\nu\xi}(2)\rangle, \label{Eq-charge-susceptibility}
\end{align}
where generalized density operators are defined by
\begin{align}
n^{}_{\alpha\beta}(1) &:= n^{}_{\alpha\beta,\uparrow}(1)+n^{}_{\alpha\beta,\downarrow}(1)\\
n^{}_{\alpha\beta,\sigma}(1) &:= c^\dagger_{\beta,\sigma}(1^+)c^{}_{\alpha,\sigma}(1)
\end{align}
and it was shown that certain sum rules that are called \textit{local charge sum rules} can be defined in the paramagnetic phase,
\begin{align}
\frac{1}{\tilde\beta N_{\vec q}}\sum_{q} \chi^{ch}_{\mu\mu\mu\mu}(q) =& \langle n^{}_{\mu}\rangle + 2\langle n^{}_{\mu,\uparrow}n^{}_{\mu,\downarrow}\rangle - \langle n_\mu\rangle^2,\\
\frac{1}{\tilde\beta N_{\vec q}}\sum_{q} \chi^{ch}_{\mu\mu\nu\nu}(q) =& 2\langle n_{\mu,\uparrow}n_{\nu,\uparrow}\rangle + 2\langle n_{\mu,\uparrow}n_{\nu,\downarrow}\rangle - \langle n_\mu\rangle \langle n_\nu\rangle
\end{align}
and for $\mu\neq\nu$,
\begin{align}
&\frac{1}{\tilde\beta N_{\vec q}}\sum_{q} \mathrm{Re}\chi^{ch}_{\mu\nu\mu\nu}(q) \nonumber\\
=&  \frac{1}{2}\langle n_\mu +n_\nu\rangle - 4\langle n_{\mu,\uparrow}n_{\nu,\uparrow}\rangle
+ 2\langle n_{\mu,\uparrow}n_{\nu,\downarrow}\rangle -\langle n_{\nu\mu} \rangle \langle n_{\mu\nu}\rangle,\label{Eq: charge_sumrules}
\end{align}
where the expectation values can be calculated from the Green's function: 
\begin{align}
\left\langle n_{\alpha\beta,\sigma}\right\rangle&= \lim_{\tau\rightarrow 0^-} G_{\alpha\beta,\sigma}(0,\tau) \nonumber\\
&=\frac{1}{\tilde \beta N_{\vec k}}\lim_{\tau\rightarrow 0^-}\sum_{\vec k} \sum_{n\in\mathbb{Z}}e^{-i\omega_n\tau}G_{\alpha\beta,\sigma}(\vec k,i\omega_n)
\end{align}
Those equalities can be directly proven from the definition (eq.~\eqref{Eq-charge-susceptibility}) and the Pauli principle. The importance of those strictly valid equations for TPSC will be explained later.
Taking a closer look at eq.~\eqref{Eq-dGdphi} one identifies
\begin{align}
-\sum_{\sigma\sigma'}\left.\frac{\delta G_{\mu\lambda,\sigma}(1,1^+)_\phi}{\delta\phi_{\nu\xi,\sigma'}(2^+,2)}\right|_{\phi=0}=\chi^{ch}_{\lambda\mu\xi\nu}(1,2) \label{Eq-dGdphi-chi}
\end{align}
which motivates the definition of a generalized \textit{three-point charge susceptibility}
\begin{align}
\chi^{ch}_{\lambda\mu\xi\nu}(1,3;2) &:= -\sum_{\sigma\sigma'}\left.\frac{\delta G_{\mu\lambda,\sigma}(1,3)_\phi}{\delta\phi_{\nu\xi,\sigma'}(2^+,2)}\right|_{\phi=0} \label{Eq-chi_ch}
\end{align}
that reproduces the previously defined charge susceptibility (eq.~\eqref{Eq-charge-susceptibility}) in the limit $3\rightarrow 1^+$.

This can be further evaluated with the self-consistent equation for $\frac{\delta G}{\delta \phi}$ (eq.~\eqref{Eq-dGdphi-sc}) and leads to
\begin{align}
&\chi^{ch}_{\lambda\mu\xi\nu}(1,3;2) \nonumber\\
=& -2 G_{\mu\nu,\sigma}(1,2^+) G_{\xi\lambda,\sigma}(2,3) \nonumber\\
&+ G_{\mu\bar\beta,\sigma}(1,\bar 4)\Gamma^{ch}_{\bar\gamma\bar\beta\bar\lambda\bar\rho}(\bar 4,\bar 5;\bar 6,\bar 7)\chi^{ch}_{\bar\lambda\bar\rho\xi\nu}(\bar 6,\bar 7;2) G_{\bar\gamma\lambda,\sigma}(\bar 5,3), \label{Eq-Bethe-Salpeter-ch}
\end{align}
where one defined the \textit{(irreducible) charge vertex} as
\begin{align}
\Gamma^{ch}_{\gamma\beta\lambda\rho}( 4, 5; 6, 7) &:= \sum_\sigma\left.\frac{\delta \Sigma_{\beta \gamma,\sigma}(4, 5)_\phi}{\delta G_{\rho\lambda,\uparrow}(6,7)_\phi}\right|_{\phi=0}.
\end{align}

Similarly, the spin susceptibilities are defined as the linear response to spin perturbations from
\begin{align}
\vec S_{\alpha\beta}(1):=&(c_{\alpha,\uparrow}^\dag(1)~~ c_{\alpha,\downarrow}^\dag(1))\cdot\vec \sigma\cdot(c^{}_{\beta,\uparrow}(1)~~c^{}_{\beta,\downarrow}(1))^T \\
S^{\pm}_{\alpha\beta}(1) :=& S^x_{\alpha\beta}(1)\pm iS^y_{\alpha\beta}(1)=\begin{cases}
c^\dag_{\alpha,\uparrow}(1)c^{}_{\beta,\downarrow}(1) &, + \\
c^\dag_{\alpha,\downarrow}(1)c^{}_{\beta,\uparrow}(1) &, -
\end{cases},
\end{align}
where $\vec \sigma = (\sigma_x,\sigma_y,\sigma_z)^{T}$ denotes the vector of Pauli spin matrices.
The spin susceptibility is then defined as the linear response
\begin{align}
\chi^{sp}_{\lambda\mu\nu\xi}(1,2):=& 4\langle T_\tau S^z_{\lambda\mu}(1)S^z_{\xi\nu}(2)\rangle - 4\langle S^z_{\lambda\mu}(1)\rangle\langle S^z_{\xi\nu}(2)\rangle \nonumber\\
=&  2\langle T_\tau c^\dag_{\lambda,\uparrow}(1)c^{}_{\mu,\uparrow}(1)c^\dagger_{\xi,\uparrow}(2)c^{}_{\nu,\uparrow}(2)\rangle\nonumber\\
&-2\langle  T_{\tau}c^\dag_{\lambda,\downarrow}(1)c^{}_{\mu,\downarrow}(1)c^\dagger_{\xi,\uparrow}(2)c^{}_{\nu,\uparrow}(2)\rangle \label{Eq:chi_sp_operators}
\end{align}
and
\begin{align}
\chi^{sp,\pm}_{\lambda\mu\nu\xi}(1,2):=& 4\langle T_\tau S^+_{\lambda\mu}(1)S^-_{\xi\nu}(2)\rangle = 2\chi^{sp}_{\lambda\mu\nu\xi}(1,2).\label{Eq: chi_sp_pm}
\end{align}
This definitions allow together with the Pauli principle and the fluctuation-dissipation theorem already to find important equalities
\begin{align}
\frac{1}{\tilde\beta N_{\vec q}}\sum_{q} \chi^{sp}_{\alpha\alpha\alpha\alpha}(q) =& 2\langle n^{}_{\alpha,\uparrow}\rangle - 2\langle n^{}_{\alpha,\uparrow}n^{}_{\alpha,\downarrow}\rangle \label{Eq: spinsumrule1}\\
\frac{1}{\tilde\beta N_{\vec q}}\sum_q \chi^{sp}_{\alpha\alpha\beta\beta}(q) =& 2 \langle n^{}_{\alpha,\uparrow}n^{}_{\beta,\uparrow}\rangle - 2\langle n^{}_{\alpha,\uparrow}n^{}_{\beta,\downarrow}\rangle,\label{Eq: spinsumrule2}\\
\frac{1}{\tilde\beta N_{\vec q}}\sum_q \mathrm{Re}\chi^{sp}_{\alpha\beta\alpha\beta}(q) =&  \langle n^{}_{\alpha,\uparrow}\rangle + \langle n^{}_{\beta,\uparrow}\rangle - 2\langle n^{}_{\alpha,\uparrow}n^{}_{\beta,\downarrow}\rangle\label{Eq: spinsumrule3}
\end{align}
that hold strictly and are called \textit{local spin sum rules}.
The generalized spin susceptiblity is defined as
\begin{align}
&\chi^{sp}_{\lambda\mu\xi\nu}(1,3;2) \nonumber\\
:=& -\sum_{\sigma\sigma'}\sigma\sigma'\left.\frac{\delta G_{\mu\lambda,\sigma}(1,3)_\phi}{\delta\phi_{\nu\xi,\sigma'}(2^+,2)}\right|_{\phi=0} \nonumber\\
=&-2G_{\mu\nu,\sigma}(1,2) G_{\xi\lambda,\sigma}(2,3)\nonumber\\
&- G_{\mu\bar\beta,\sigma}(1,\bar 4)\Gamma^{sp}_{\bar\gamma\bar\beta\bar\lambda\bar\rho}(\bar 4,\bar 5;\bar 6,\bar 7)\chi^{sp}_{\bar\lambda\bar\rho\xi\nu}(\bar 6,\bar 7;2) G_{\bar\gamma\lambda,\sigma}(\bar 5,3), \label{Eq-Bethe-Salpeter-sp}
\end{align}
where we adapt the common notation $\uparrow\equiv 1,~\downarrow\equiv -1$ and use the definition of the \textit{(irreducible) spin vertex}
\begin{align}
\Gamma^{sp}_{\gamma\beta\lambda\rho}(4,5;6,7) &:=\sum_{\sigma}\sigma\left.\frac{\delta \Sigma_{\beta \gamma,\sigma}(4,5)_\phi}{\delta G_{\rho\lambda,\downarrow}(6,7)_\phi}\right|_{\phi=0}.\label{Eq-gamma-sp}
\end{align}

\clearpage
\newpage
\subsection{Self-energy and susceptibilities}

Now, all equations can be put together to derive the self-energy as a function of the generalized three-point spin and charge susceptibilities.
Starting from the implicit equation for the self-energy (eq.~\eqref{Eq-self-energy-dGdphi}) 
and multiplying with $\left(G^{-1}\right)_{\bar\gamma\delta,\sigma}(\bar2,2)$ results in
\begin{widetext}
	\begin{align}
	\Sigma_{\alpha\delta,\sigma}(1,2) &=-\sum_{\beta}U_{\beta\alpha}\left[-\frac{1}{4}\left(\chi^{ch}_{\bar\gamma\alpha\beta\beta}(1,\bar 3;1)-\chi^{sp}_{\bar\gamma\alpha\beta\beta}(1,\bar 3;1)\right)\left(G^{-1}\right)_{\bar\gamma\delta,\sigma}(\bar 3,2)-n_{\beta,-\sigma}(1)\delta_{\alpha,\delta}\delta(1-2)\right]\nonumber\\
	&-\sum_{\substack{\beta \\ \beta\neq\alpha}}(U_{\beta\alpha}-J_{\beta\alpha})\left[-\frac{1}{4}\left(\chi^{ch}_{\bar\gamma\alpha\beta\beta}(1,\bar 3;1)+\chi^{sp}_{\bar\gamma\alpha\beta\beta}(1,\bar 3;1)\right)\left(G^{-1}\right)_{\bar\gamma\delta,\sigma}(\bar 3,2)-n_{\beta,\sigma}(1)\delta_{\alpha,\delta}\delta(1-2)\right]\nonumber\\
	&-\sum_{\substack{\beta\\\beta\neq\alpha}}J_{\beta\alpha}\left[-\frac{1}{4}\left(\chi^{ch}_{\bar\gamma\beta\alpha\beta}(1,\bar 3;1)-\chi^{sp}_{\bar\gamma\beta\alpha\beta}(1,\bar 3;1)\right)\left(G^{-1}\right)_{\bar\gamma\delta,\sigma}(\bar 3,2) -n_{\alpha\beta,-\sigma}(1)\delta_{\beta,\delta}\delta(1-2)\right.\nonumber\\
	&\hspace{1.5cm}+\left.-\frac{1}{4}\left(\chi^{ch}_{\bar\gamma\beta\beta\alpha}(1,\bar 3;1)-\chi^{sp}_{\bar\gamma\beta\beta\alpha}(1,\bar 3;1)\right)\left(G^{-1}\right)_{\bar\gamma\delta,\sigma}(\bar 3,2)-n_{\beta\alpha,-\sigma}(1)\delta_{\beta,\gamma}\delta(1-2)\right].
	\end{align}
\end{widetext}
Making use of the equations of motion for the three-point susceptibilities (eqs.~\eqref{Eq-Bethe-Salpeter-sp},~\eqref{Eq-Bethe-Salpeter-ch}) and dropping the $\phi$ subscript to reduce notation yields to an expression of the self-energy in terms of three-point susceptibilities, four point irreducible spin and charge vertices and full Green's functions, namely
\begin{widetext}
	\begin{align}
	&\Sigma_{\alpha\delta,\sigma}(1,2) \nonumber\\
	=&
	\left[\sum_{\beta}U_{\beta\alpha}n_{\beta,-\sigma}(1)\delta_{\alpha,\delta}+\sum_{\beta\neq\alpha}(U_{\beta\alpha}-J_{\beta\alpha})\left(n_{\beta,\sigma}(1)\delta_{\alpha,\delta}-n_{\alpha\delta,\sigma}(1)\delta_{\beta,\delta}\right)+J_{\delta\alpha}(n_{\alpha\delta,-\sigma}(1)+n_{\delta\alpha,-\sigma}(1))(1-\delta_{\alpha,\delta}))\right]\delta(1-2)\nonumber\\
	&+ \frac{1}{4}\sum_{\gamma,\beta}G_{\gamma\beta}(1,\bar 3)\left(\Gamma^{ch}_{\delta\beta\bar{\epsilon}\bar{\delta}}(\bar 3,2;\bar 7,\bar 8)\chi^{ch}_{\bar{\epsilon}\bar{\delta}\bar\lambda\bar\rho}(\bar 7,\bar 8;1)\Gamma^{ch,0}_{\bar\lambda\bar\rho\alpha\gamma} +\Gamma^{sp}_{\delta\beta\bar{\epsilon}\bar{\delta}}(\bar 3,2;\bar 7,\bar 8)\chi^{sp}_{\bar{\epsilon}\bar{\delta}\bar\lambda\bar\rho}(\bar 7,\bar 8;1)\Gamma^{sp,0}_{\bar\lambda\bar\rho\alpha\gamma}\right),\label{Eq-self-energy-4point}
	\end{align}
\end{widetext}
where one defines
\begin{align}
\Gamma^{ch,0}_{\alpha\beta\gamma\delta}=\begin{cases}
U_{\alpha\alpha}   & \alpha=\beta=\gamma=\delta\\
2U_{\alpha\gamma}-J_{\alpha\gamma} & \alpha=\beta\neq\gamma=\delta\\
J_{\alpha\beta}   & \alpha=\gamma\neq\beta=\delta\\
J_{\alpha\beta}  & \alpha=\delta\neq\beta=\gamma\\
0   & else
\end{cases}
\end{align}
and
\begin{align}
\Gamma^{sp,0}_{\alpha\beta\gamma\delta}=\begin{cases}
U_{\alpha\alpha}   & \alpha=\beta=\gamma=\delta\\
J_{\alpha\gamma}  & \alpha=\beta\neq\gamma=\delta\\
J_{\alpha\beta}   & \alpha=\gamma\neq\beta=\delta\\
J_{\alpha\beta}  & \alpha=\delta\neq\beta=\gamma\\
0   & else
\end{cases}.\label{Eq: Gamma_sp0}
\end{align}
\subsubsection{The vertices $\Gamma^{sp/ch,0}$: Differences between RPA and TPSC}
Comparing the interaction vertices with RPA results~\onlinecite{Mochizuki2005} one observes a difference in the non-interacting vertices
\begin{align}
\Gamma^{ch,0,RPA}_{\alpha\beta\gamma\delta}=\begin{cases}
U_{\alpha\alpha}   & \alpha=\beta=\gamma=\delta\\
2U_{\alpha\gamma}-J_{\alpha\gamma} & \alpha=\beta\neq\gamma=\delta\\
-U_{\alpha\beta}+2J_{\alpha\beta}   & \alpha=\gamma\neq\beta=\delta\\
J_{\alpha\beta}  & \alpha=\delta\neq\beta=\gamma\\
0   & else
\end{cases}\label{Eq: Gamma_ch_RPA}
\end{align}
and 
\begin{align}
\Gamma^{sp,0,RPA}_{\alpha\beta\gamma\delta}=\begin{cases}
U_{\alpha\alpha}   & \alpha=\beta=\gamma=\delta\\
J_{\alpha\gamma}  & \alpha=\beta\neq\gamma=\delta\\
U_{\alpha\beta}   & \alpha=\gamma\neq\beta=\delta\\
J_{\alpha\beta}  & \alpha=\delta\neq\beta=\gamma\\
0   & else
\end{cases}.\label{Eq: Gamma_sp_RPA}
\end{align}
This is because in RPA the spin and charge vertices $\frac{\delta \Sigma}{\delta G}$ are calculated while discarding every higher order contribution in eq.~\eqref{Eq-self-energy-dGdphi}, i.e. the RPA vertices are constructed from functionally differentiating the Hartree-Fock self-energy.

\subsection{The multi-orbital Two-Particle Self-Consistent approach}
\label{sec: TPSC eq}
So far, all expressions derived are exact and the approximations to determine $\Sigma$ and $\chi^{sp/ch}$ are made in this section.
Analogous to the single-orbital TPSC~\onlinecite{VilkTremblay1997} and following the functional derivative formalism~\onlinecite{LuttingerWard1960,Luttinger1960} it is commonly assumed that the Legendre transform exists for the free energy $F[\phi]=-T\ln Z[\phi]$ to generate the so-called Kadanoff-Baym functional $\Omega[G]=F[\phi]-\text{tr}(\phi G)$. The explicit form of the Kadanoff-Baym functional can be constructed from the Dyson equation and one encounters the so-called Luttinger Ward functional $\Phi[G]$ which is a functional of the fully dressed single-particle Green's function $G$.
It is the sum of all closed two-particle irreducible skeleton diagrams that can be constructed from $G$ and the on-site interactions $U,~J$.
The first and second order functional derivatives of $\Phi[G]$ are related to
the single-particle self-energy $\Sigma$ and the irreducible vertex function $\Gamma$, respectively
\begin{align}
\Sigma 
&= \frac{\delta }{\delta G} \Phi[G], \\
\Gamma 
&= \frac{\delta \Sigma}{\delta G} 
= \frac{\delta^2 }{\delta G^2 } \Phi[G] .
\end{align}
Any approximation to the Luttinger-Ward functional that consists of 
a diagrammatic truncation by taking only a certain convergent subset 
of diagrams into account can be shown to be conserving in the Baym-Kadanoff sense.
The most simple example is the Hartree-Fock approximation, which 
only includes the first-order diagrams contributing to the Luttinger-Ward functional 
(see fig.~\ref{fig: diag} top line).
The well-known FLEX approximation consists of taking a subset of diagrams that can be summed via the geometrical series~\onlinecite{Bickers1989,BickersPRB1991}.
Another widely used approximation is the dynamical mean-field approximation,
which approximates the Luttinger-Ward functional by taking into account only
diagrams generated from local propagators $\Phi[G] \approx \Phi[G_{loc}]$
summed up to all orders by solving a local impurity model.

In the multi-orbital TPSC formulation of~\onlinecite{Miyahara2013,Zantout2019} one proceeds as in the single-band TPSC~\onlinecite{VilkTremblay1997} approach where the Luttinger-Ward functional is approximated by $\Phi=G\Gamma G$ where $\Gamma$ is static and local. In the same fashion as in the single-orbital case this leads to a constant self-energy $\Sigma(1,2)=\Gamma n(1)\delta(1-2)$ and local and static (but orbital-dependent) spin and charge vertices
\begin{align}
\Gamma^{sp/ch}_{\alpha\beta\gamma\delta}(1,2;3,4) 
&= \Gamma^{sp/ch}_{\alpha\beta\gamma\delta} \delta(1-3)\delta(2-4)\delta(1-2).\label{Eq-vertex_const}
\end{align}
The TPSC self-energy is closely related to the Hartree-Fock self-energy but with a renormalized effective interaction $\Gamma$. This is very different to the Hartree-Fock approximation, which equates $\Gamma$ with the bare interactions $U,~J$ while in TPSC one \textit{a priori} does not impose a limitation on the value of $\Gamma$.\\
The constant self-energy contribution can be assumed to be already included in the input from density functional theory.
To further improve the self-energy within TPSC one reinserts the bare Green's function into the self-energy equation eq. \eqref{Eq-self-energy-4point}. This gives an improved self-energy where the collective modes enter while keeping the level of appoximation, i.e. susceptibilities are computed from $G^0$ and the input Green's function for the self-energy is also $G^0$. This can be motivated by the fact that collective modes influence single-particle properties but the opposite effect is much smaller.\\
Obviously, as argued in the single-band TPSC~\onlinecite{VilkTremblay1997} one has to keep all equations at this level of iteration, i.e. single-shot calculations, since reiterating the susceptibilities or self-energy with the full Green's function would not only violate the local spin and charge sum rules but also be in contradiction to the assumption of static and local irreducible vertices $\Gamma^{sp/ch}$. 
Since in~\onlinecite{Miyahara2013,Zantout2019} one uses input from DFT it is assumed that the static Hartree-Fock terms are already accounted for and therefore one drops them in the self-energy expression of eq.~\eqref{Eq-self-energy-4point}.
Thus, equation~\eqref{Eq-self-energy-4point}) and equation~\eqref{Eq-vertex_const} lead to the final expression for the self-energy,
\begin{align}
&\Sigma_{\alpha\delta,\sigma}(1,2)\nonumber= \frac{1}{4}\sum_{\gamma,\beta}G^0_{\gamma\beta}(1,2)\times\nonumber\\
&\times\left(\Gamma^{ch}_{\delta\beta\bar{\epsilon}\bar{\delta}}\chi^{ch}_{\bar{\epsilon}\bar{\delta}\bar\lambda\bar\rho}(2,1)\Gamma^{ch,0}_{\bar\lambda\bar\rho\alpha\gamma}+\Gamma^{sp}_{\delta\beta\bar{\epsilon}\bar{\delta}}\chi^{sp}_{\bar{\epsilon}\bar{\delta}\bar\lambda\bar\rho}(2,1)\Gamma^{sp,0}_{\bar\lambda\bar\rho\alpha\gamma}\right).
\end{align}
A simple Fourier transformation leads to
\begin{align}
&\Sigma_{\alpha\delta,\sigma}(k)=\frac{T}{4N_{\vec{q}}}\sum_{q}\left[\Gamma^{ch}\chi^{ch}\Gamma^{ch,0}+\Gamma^{sp}\chi^{sp}\Gamma^{sp,0}\right]_{\delta\bar{\beta}\alpha\bar{\gamma}}(q)\times\nonumber\\
&\times G^0_{\bar\gamma\bar\beta}(k+q).\label{eq: Sigma_final}
\end{align}
The resulting self-energy has the shape from paramagnon theories and allows for interpretations where the coupling to bosonic modes in the spin and charge channel is dressed with renormalized vertices due to higher order correlation functions.
Since the derivation of the self-energy was done only in the longitudinal channel the four-point vertex $\Gamma$ does not fulfill crossing symmetry (see also Sec.~\ref{sec: Outlook}).\\
One continues with the susceptibilities (eqs.~\eqref{Eq-Bethe-Salpeter-sp}, \eqref{Eq-Bethe-Salpeter-ch}) by inserting the TPSC spin and charge vertices from eq.~ \eqref{Eq-vertex_const} which simplifies the expressions to
\begin{align}
\chi^{sp}_{\lambda\mu\xi\nu}(1,1^+;2) =& 2\chi^0_{\lambda\mu\xi\nu}(1,2) + \Gamma_{\bar{\gamma}\bar{\beta}\bar{\lambda}\bar{\rho}}^{sp}\chi_{\bar{\lambda}\bar{\rho}\xi\nu}^{sp}(\bar{3},2)\chi_{\lambda\mu\bar{\gamma}\bar{\beta}}^{0}(1,\bar{3})
\end{align}
and thus
\begin{align}
2\chi_{\lambda\mu\xi\nu}^0(1,2) =&  \left(\delta_{\lambda\bar{\lambda}}\delta_{\mu\bar{\rho}}-\chi_{\lambda\mu\bar{\gamma}\bar{\beta}}^0(1,\bar{3})\Gamma^{sp}_{\bar{\gamma}\bar{\beta}\bar{\lambda}\bar{\rho}}\right)\chi^{sp}_{\bar\lambda\bar{\rho}\xi\nu}(\bar{3},2).
\end{align}
By taking advantage of the index combination $(\alpha\beta\gamma\delta)\rightarrow(\alpha\beta),(\gamma,\delta)$ to reduce tensor equations to matrix equations it was shown that a Fourier transformation yields
\begin{align}
\chi^{sp}_{\lambda\mu\xi\nu}(q) &= [1-\chi^0\Gamma_{sp}]_{\lambda\mu\bar\lambda\bar{\rho}}^{-1}2\chi^0_{\bar\lambda\bar{\rho}\xi\nu}(q).
\end{align}
Analogously, the equation in the charge channel is
\begin{align}
\chi^{ch}_{\lambda\mu\xi\nu}(q) &= [1+\chi^0\Gamma_{ch}]_{\lambda\mu\bar\lambda\bar{\rho}}^{-1}2\chi^0_{\bar\lambda\bar{\rho}\xi\nu}(q).
\end{align}

The high-frequency behavior of the self-energy can be calculated via the local spin and charge sum rules but does not coincide with the exact high frequency result\onlinecite{Gull2011,Wang2011}.
The reason lies within the constant vertices $\Gamma^{sp/ch}$ since a proper tail in the longitudinal particle-hole channel can only occur if the interaction vertices renormalize to the bare interaction in the high-frequency limit\onlinecite{VilkTremblay1997}.
The same statement is not true for the transversal particle-hole channel although averaging the TPSC self-energy expressions gave improved high-frequency behavior~\onlinecite{Moukouri2000} which leads us to the assumption that this improvement is an error cancellation effect.

\subsection{Ansatz equations for the irreducible vertices}
So far, we only described how a local and static four-point vertex can simplify the self-energy and susceptibility expressions.
In this subsection it is explained how to determine its value.
\subsubsection{The spin vertex}
In order to determine the renormalized vertices $\Gamma^{sp}$ within TPSC one makes use of the local spin sum rules (eqs.~\eqref{Eq: spinsumrule1}, \eqref{Eq: spinsumrule2} and \ref{Eq: spinsumrule3}). 
Unfortunately, these sum rules also include unknowns, namely the double occupations $\langle n_{\mu,\sigma}n_{\nu,\sigma'}\rangle$ and the system of equations is therefore under-determined. 
Therefore, more information is needed to fix both the double occupations and the vertex $\Gamma^{sp}$.\\
The simplest ansatz is to do a Hartree-Fock decoupling for the right-hand side of $\Sigma G$ (eq.~\eqref{Eq-self-energy}) for each expectation value and write a prefactor $A,B$ in front to recover the result for equal time/position/orbital evaluation, i.e. $2 \rightarrow 1$ and $\alpha=\gamma$:
\begin{align}
&\Sigma_{\alpha\bar\beta,\sigma}(1,\bar 3) G_{\bar \beta \gamma,\sigma}(\bar 3, 2) \nonumber\\
=&-\sum_{\beta}U_{\beta\alpha} \left\langle n_{\beta,-\sigma}(1^+)c_{\alpha,\sigma}(1)c^\dag_{\gamma,\sigma}(2)\right\rangle\nonumber\\
&-\sum_{\substack{\beta \\ \beta\neq\alpha}}\frac{1}{2}(U_{\beta\alpha}-J_{\beta\alpha}) \left\langle n_{\beta,\sigma}(1^+)c_{\alpha,\sigma}(1)c^\dag_{\gamma,\sigma}(2)\right\rangle\nonumber\\
&+\sum_{\substack{\beta\\\beta\neq\alpha}}J_{\beta\alpha}\left(\left\langle c^\dag_{\beta,-\sigma}(1^+)c_{\beta,\sigma}(1)c_{\alpha,-\sigma}(1)c^\dag_{\gamma,\sigma}(2)\right\rangle\right.\nonumber\\
&~~~~~~~~~~~~~+\left.\left\langle c^\dagger_{\alpha,-\sigma}(1^{++})c_{\beta,\sigma}(1^+)c_{\beta,-\sigma}(1)c^\dag_{\gamma,\sigma}(2)\right\rangle\right)\nonumber\\
\stackrel{HF}{\approx}&~ A^\sigma_{\alpha} n_{\alpha,-\sigma}(1) G_{\alpha\gamma,\sigma}(1,2)\nonumber\\
&+\sum_{\substack{\beta,\sigma' \\ \beta\neq\alpha}} B^{\sigma'\sigma}_{\beta\alpha} n_{\beta,\sigma'}(1) G_{\alpha\gamma,\sigma}(1,2).\label{Eq: AnsatzGamma}
\end{align}
To recover now the original result for equal time/position we set
\begin{align}
A^{\sigma}_{\alpha} &= U_{\alpha\alpha}\frac{\langle n_{\alpha,\sigma}n_{\alpha,-\sigma}\rangle}{\langle n_{\alpha,\sigma}\rangle\langle n_{\alpha,-\sigma}\rangle}\nonumber\\
B^{\sigma\sigma}_{\beta\alpha} &= 
\left(U_{\beta\alpha}-J_{\beta\alpha}\right)\frac{\langle n_{\beta,\sigma}n_{\alpha,\sigma}\rangle}{\langle n_{\beta,\sigma}\rangle\langle n_{\alpha,\sigma}\rangle}~~,\beta\neq\alpha\nonumber\\
B^{\sigma-\sigma}_{\beta\alpha} &= U_{\beta\alpha}\frac{\langle n_{\beta,\sigma}n_{\alpha,-\sigma}\rangle}{\langle n_{\beta,\sigma}\rangle\langle n_{\alpha,-\sigma}\rangle}~~,\beta\neq\alpha.\label{Eq-Usp_ansatz_pre}
\end{align}
Substituting $A,~B$ back into the ansatz equation (eq.~\eqref{Eq: AnsatzGamma}) and multiplying with $\left(G^{-1}\right)_{\bar{\gamma}\nu}(\bar2,2)$ one obtains
\begin{align}
\Sigma_{\alpha\nu,\sigma}(1,2) &\approx A^{\sigma}_{\nu} n_{\alpha,-\sigma}(1) \delta_{\alpha,\nu}\delta(1-2)\nonumber\\
&+\sum_{\substack{\beta,\sigma' \\ \beta\neq\nu}} B^{\sigma'\sigma}_{\beta\nu} n_{\beta,\sigma'}(1) \delta_{\alpha,\nu}\delta(1-2).
\end{align} 
To get now the renormalized vertices one performs the functional derivatives from eq.~\eqref{Eq-gamma-sp}. This leads to the ansatz
\begin{align}
&\Gamma^{sp}_{\gamma\beta\lambda\rho}(1,2;3,4)\nonumber\\ 
&= A^\sigma_{\beta} \delta_{\beta,\rho}\delta_{\beta,\lambda}\delta_{\beta,\gamma}\delta(1-3)\delta(1^+-4)\delta(1-2)\nonumber\\
&+ B^{\uparrow\downarrow}_{\rho\beta} (1-\delta_{\rho,\beta})\delta_{\lambda,\rho}\delta_{\beta,\gamma}\delta(1-3)\delta(1^+-4)\delta(1-2)\nonumber\\
&- B^{\downarrow\downarrow}_{\rho\beta} (1-\delta_{\rho,\beta})\delta_{\lambda,\rho}\delta_{\beta,\gamma}\delta(1-3)\delta(1^+-4)\delta(1-2).\label{Eq: Gamma_local}
\end{align}
Thus, equation~\eqref{Eq: Gamma_local} together with the local spin sum rules gives us a set of equations to uniquely determine the spin vertex $\Gamma^{sp}$. 
Moreover, in~\onlinecite{Zantout2019} one further sets $\Gamma^{sp}_{\mu\nu\nu\mu}=\Gamma^{sp}_{\mu\nu\mu\nu}=\Gamma^{sp}_{\mu\mu\nu\nu}$ due to the symmetry of $\Gamma^{sp,0}$ (eq.~\eqref{Eq: Gamma_sp0}).
This is in contrast to~\onlinecite{Miyahara2013} which is discussed in sec.~\ref{sec: comparison}.\\ 
Since the ansatz equations in \eqref{Eq-Usp_ansatz_pre} are not particle-hole symmetric it was shown in Ref.~\onlinecite{Zantout2019} that this can be enforced by symmetrizing those expressions,
\begin{align}
A^{\sigma}_{\mu} =& U_{\mu\mu}\frac{1}{2}\left(\frac{\langle n_{\mu,\sigma}n_{\mu,-\sigma}\rangle}{\langle n_{\mu,\sigma}\rangle\langle n_{\mu,-\sigma}\rangle} +\right.\nonumber\\
&\left.\hspace{1cm}+ \frac{\langle (1-n_{\mu,\sigma})(1-n_{\mu,-\sigma})\rangle}{\langle 1-n_{\mu,\sigma}\rangle\langle 1-n_{\mu,-\sigma}\rangle}\right)\nonumber\\
B^{\sigma\sigma}_{\alpha\mu} &\stackrel{\alpha\neq\mu}{=} 
\left(U_{\alpha\mu}-J_{\alpha\mu}\right)\frac{1}{2}\left(\frac{\langle n_{\alpha,\sigma}n_{\mu,\sigma}\rangle}{\langle n_{\alpha,\sigma}\rangle\langle n_{\mu,\sigma}\rangle} +\right.\nonumber\\
&\left.\hspace{2.8cm}+ \frac{\langle (1-n_{\alpha,\sigma})(1-n_{\mu,\sigma})\rangle}{\langle 1-n_{\alpha,\sigma}\rangle\langle 1-n_{\mu,\sigma}\rangle}\right)\nonumber\\
B^{\sigma-\sigma}_{\alpha\mu} &\stackrel{\alpha\neq\mu}{=} U_{\alpha\mu}\frac 1 2
\left(\frac{\langle n_{\alpha,\sigma}n_{\mu,-\sigma}\rangle}{\langle n_{\alpha,\sigma}\rangle\langle n_{\mu,-\sigma}\rangle} +\right.\nonumber\\
&\left.\hspace{1.6cm}+ \frac{\langle (1-n_{\alpha,\sigma})(1-n_{\mu,-\sigma})\rangle}{\langle 1-n_{\alpha,\sigma}\rangle\langle 1-n_{\mu,-\sigma}\rangle}\right).\label{Eq-Usp_ansatz}
\end{align}
This enforcement of particle-hole symmetry is only one way to deal with the breaking of particle-hole symmetry in eq.~\eqref{Eq: Gamma_local}. 
Alternatively, one can do a particle-hole transformation in the case of electron doping and keep eq.~\eqref{Eq: Gamma_local} in the case of hole doping as explained in~\onlinecite{VilkTremblay1997}.\\
Note, that the ansatz fails for $J=0,~U\neq 0$ since in that case the ansatz for $\Gamma^{sp}_{\mu\mu\nu\nu}$ renormalizes to zero and one ends up with a non-interacting double-occupancy $\langle n_{\mu}n_{\nu}\rangle=\langle n_{\mu}\rangle \langle n_{\nu}\rangle$ for all values of $U$.
\\~\\
It was shown in the original single-band TPSC that such an ansatz can be also motivated from the local-field approach for the electron gas and reproduces Kanamori-Bruecker screening~\onlinecite{Vilk1994}.\\
In principle, one would have to use the occupations $\langle n_{\alpha,\sigma}\rangle$ from the interacting system in eq.~\eqref{Eq-Usp_ansatz} but in Ref.~\onlinecite{Zantout2019} one uses the occupations of the non-interacting system and assumes that those are close to the occupations of the interacting system.
For the specific material study in Ref.~\onlinecite{Zantout2019} one can show that this is a rather good approximation and consistent with the idea of single-band TPSC that spin and charge fluctuations are effectively calculated from the bare Green's function.
This additional approximation is employed in order to avoid an additional self-consistency loop between the spin and charge susceptibilities and the interacting Green's function.\\ 
Alternatively, one could use $\langle n_{\mu,\sigma}n_{\nu,\sigma'}\rangle$ as an input from some other method like DMFT and solve the equations of the spin and charge sum rules directly (eq.~\eqref{Eq: charge_sumrules},\eqref{Eq: spinsumrule2},\eqref{Eq: spinsumrule3}) without the need of the ansatz in eq.~\eqref{Eq: Gamma_local} and \eqref{Eq-Usp_ansatz}.

\subsubsection{The charge vertex}
In addition to the remarks on the self-energy tail at the end of section~\ref{sec: TPSC eq} the value of $\Gamma^{ch}_{\mu\mu\nu\nu},~\mu\neq\nu,$ might be negative in TPSC if one would enforce the local charge sum rules without further constraints, since it allows for positive $\Im\Sigma(i\omega_n)$ which is unphysical and gives e.g. negative spectral weight.

To avoid this one restricts the elements $\Gamma^{ch}_{\mu\mu\nu\nu},\mu\neq\nu$ to be non-negative.
In Ref.~\onlinecite{Zantout2019} one optimizes the charge sum rules (eq.~\eqref{Eq: charge_sumrules}) by minimizing the difference between right-hand and left-hand side while keeping the charge vertex $\Gamma^{ch}$ positive while in Ref.~\onlinecite{Miyahara2013} the negative contributions of the charge vertex are set to zero.
Both schemes yield small errors in the local charge sum rules as is discussed in section \ref{sec: application}.

\subsection{Flow diagram of multi-orbital TPSC}
\begin{figure*}
	\includegraphics[width=\textwidth]{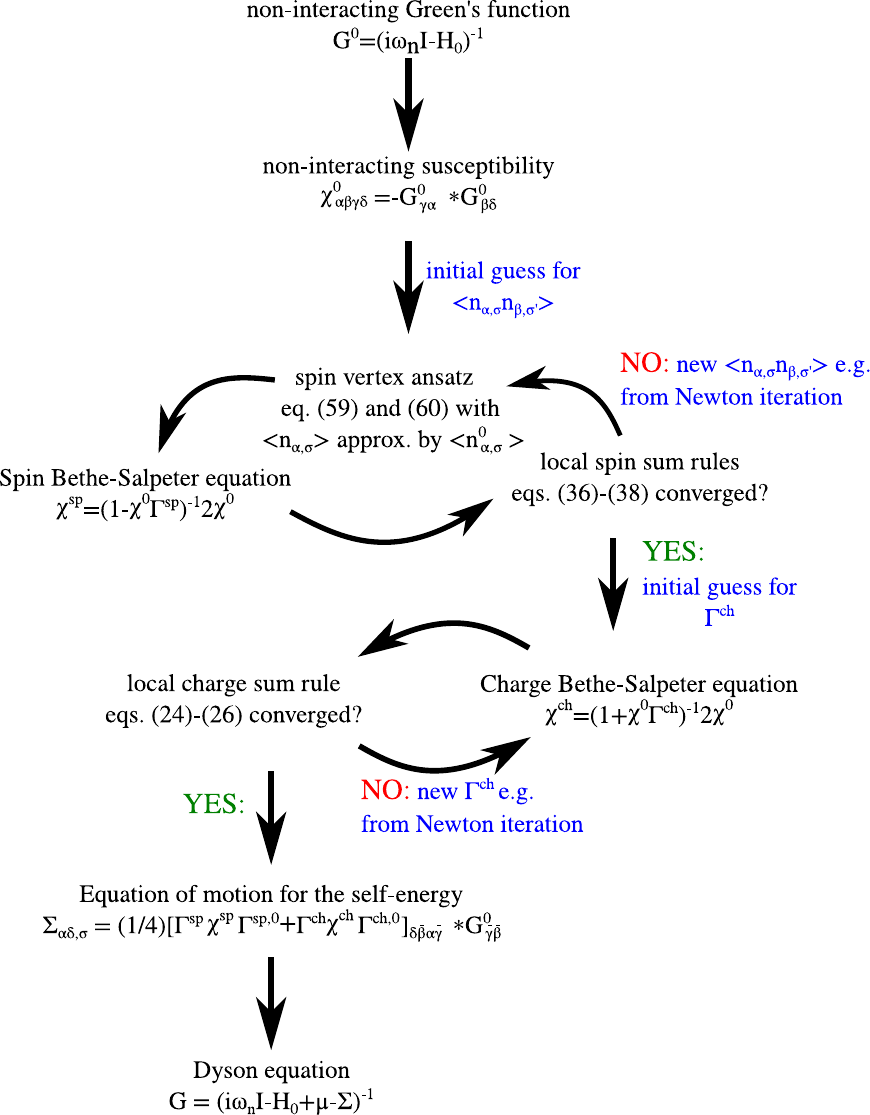}%
	\caption{\label{fig: TPSC_flow} 
		Flow diagram of the multi-orbital TPSC approach presented in this article.}
\end{figure*}

For a better overview we present in fig.~\ref{fig: TPSC_flow} a flow diagram of the multi-orbital TPSC scheme as
introduced in Ref.~\onlinecite{Zantout2019}.

\subsection{Internal accuracy check}
\label{Sec: Internal_accuracy}
The TPSC approach provides an internal accuracy check by combining the equation that relates the product of the self-energy and Green's function (eq.~\eqref{Eq-self-energy}
) to correlation functions that can be obtained from the local spin and charge sum rules.
Evaluating the left-hand side of eq.~\eqref{Eq-self-energy} at equal orbital and time/position one gets
\begin{align}
&\Sigma_{\beta\bar\alpha,\sigma}(1,\bar 2) G_{\bar\alpha\beta,\sigma}(\bar 2,1)\nonumber\\
&=\sum_{\alpha}U_{\alpha\beta}\langle n_{\alpha,-\sigma}n_{\beta,\sigma}\rangle +\sum_{\substack{\alpha\\\alpha\neq \beta}}(U_{\alpha\beta}-J_{\alpha\beta})\langle n_{\alpha,\sigma}n_{\beta,\sigma}\rangle\nonumber\\
&~-\sum_{\substack{\alpha\\\alpha\neq \beta}}J_{\alpha\beta}(\langle n_{\alpha,\sigma}n_{\beta,\sigma}\rangle-\langle n_{\alpha,-\sigma}n_{\beta,\sigma}\rangle)\nonumber\\
&~-\frac{1}{\tilde\beta N_{\vec{q}}}\sum_{\substack{q,\alpha\\\alpha\neq \beta}}\frac{J_{\alpha\beta}}{2}\left(\chi^{sp}_{\beta\alpha\alpha\beta}(q) - \chi^{ch}_{\beta\alpha\alpha\beta}(q)\right)  ,\label{Eq-tr-Sigma-G}
\end{align}
In the case of TPSC this equation is fulfilled for $G=G_0$ since the self-energy is from a single shot approach~\onlinecite{VilkTremblay1997} but in our multi-orbital case where the charge vertex cannot be calculated to fulfill the charge sum rules exactly we will also see a small deviation here (see sec.~\ref{sec: application}).

\subsection{Comparison to the multi-orbital TPSC formulation from Ref.~\cite{Miyahara2013}}
\label{sec: comparison}
While the overall form of the TPSC equations presented in \onlinecite{Zantout2019}
(and also in this review) and the ones in~\onlinecite{Miyahara2013} are exactly the
same we discuss in this section the specific
differences between both implementations and comment the implications
on the final results.\\ First, one observes that in the
formulation of Ref.~\onlinecite{Zantout2019} the charge vertex $\Gamma^{ch}$ is
determined such that its components are non-negative and minimize the error in
the local charge sum rules.
This is in contrast to the scheme in~\onlinecite{Miyahara2013}, where the negative contribution to the charge vertex component $\Gamma^{ch}_{\alpha\alpha\beta\beta}$, that eventually leads to unphysical negative spectral weight, is set to zero by hand.
Since the optimization procedure is less invasive we expect that the charge sum rules in Ref.~\onlinecite{Zantout2019} are fulfilled
to a higher degree than in Ref.~\onlinecite{Miyahara2013},
although in general the optimization procedure converges to values of $\Gamma^{ch}_{\alpha\alpha\beta\beta}$ that are very close to zero compared to the other non-trivial elements of $\Gamma^{ch}$ (see e.g. fig.~\ref{fig: Uch}).
For this reason we assume that the treatment of the charge channel is still very similar in both schemes.

Next, we find that Miyahara \textit{et al.} employ the spin vertex ansatz in eq.~\eqref{Eq-Usp_ansatz_pre} while we use the particle-hole symmetrized ansatz in eq.~\eqref{Eq-Usp_ansatz}.
It was already pointed out in~\onlinecite{Miyahara2013} that the particle-hole transformation only leads to marginal changes for the high-Tc cuprate 
superconductor studied but in general this will strongly depend on the material.
In any case, one needs to consider some kind of particle-hole symmetrization to meet the original ideas of this ansatz~\onlinecite{Singwi1968,VilkTremblay1997}.

Another important difference between both formulations is the choice of the local spin and charge sum rules that are used to determine the spin and charge vertex $\Gamma^{sp/ch}$. 
The sum rules presented here were chosen such that the right-hand side is only dependent on the double occupations and orbital occupations which allows for easier numerical root search but the outcome of both schemes should be the same.

Finally, an important difference is the form of the bare vertex functions $\Gamma^{sp/ch,0}$.
While in the derivation we reproduced here the expressions explicitly follow from the Bethe-Salpeter equation and the equation of motion of the self-energy, Miyahara \textit{et al.} employed the RPA expressions of those vertices.
This leads to a difference in the vertex components $\alpha\beta\alpha\beta$ for $\alpha\neq\beta$.\\
Since those vertices enter the self-energy expression in eq.~\eqref{eq: Sigma_final} which fulfills the self-consistency relation $\mathrm{tr}(\Sigma G^0)$ exactly (see eq.~\eqref{Eq-tr-Sigma-G}) we conclude that the formulation presented here will be more accurate in terms of this sum rule.

\section{Applications to a model system}
\label{sec: application}
In this section we present calculations on the two-orbital Hubbard model on the square lattice with nearest-neighbor hopping $t$ 
in order to show some general features of the method and point out where the limitations are.\\
The interaction matrices $U_{\mu\nu}, J_{\mu\nu}$ are in Hubbard- Kanamori form~\onlinecite{Kanamori1963} and are thus derived from the interaction values $U,J$, i.e. $U_{\mu\nu}=U$ if $\mu=\nu$ and $U_{\mu\nu}=U-2J$ else while $J_{\mu\nu}=J$.\\
All presented results are calculated at $k_BT/t=0.5$ and half filling if not mentioned differently.

\begin{figure*}
	\includegraphics[width=\textwidth]{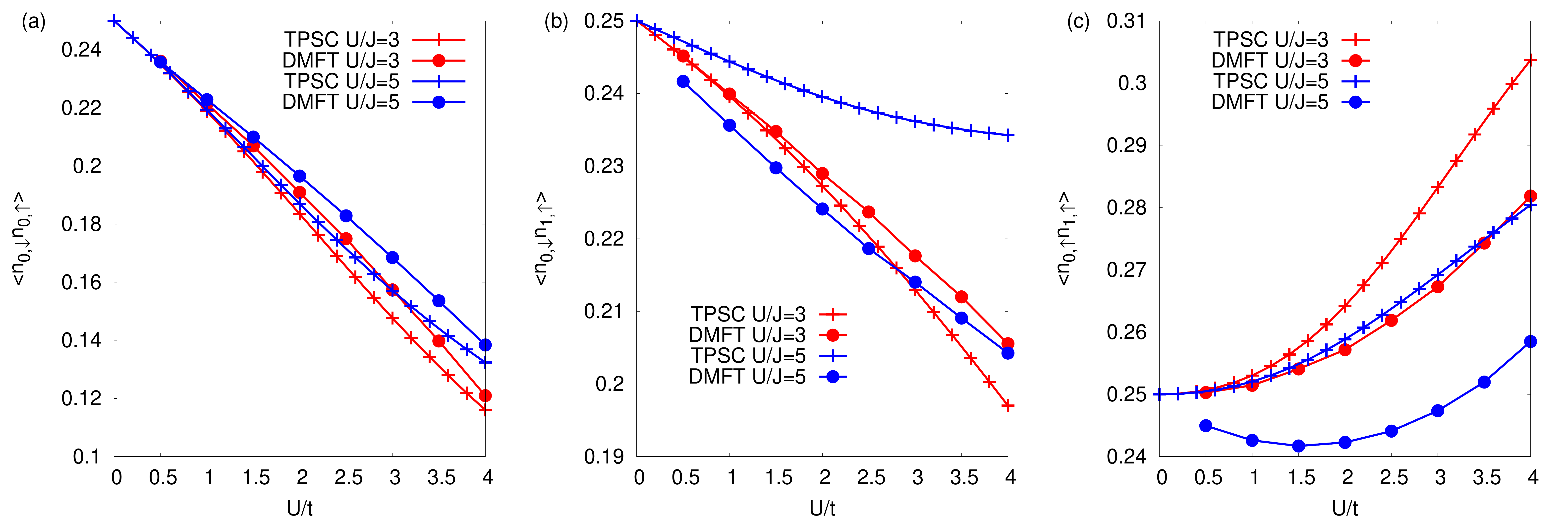}%
	\caption{\label{fig: nn}
		Double occupations $\langle n_{\mu,\sigma}n_{\nu,\sigma'}\rangle$ shown as functions of $U/t$ calculated with TPSC and DMFT. We left out redundant double occupations that can be determined either by the ones shown or by the Pauli principle. (a) The double occupations $\langle n_{\mu,\sigma}n_{\mu,-\sigma}\rangle$ show the expected behavior when increasing the interaction values $U,J$, namely they drop and this is enhanced by enlarging $J$. TPSC and DMFT results are comparable. 
		In (b) and (c) inter-orbital double occupations $\langle n_{\mu,\sigma}n_{\nu,\sigma'}\rangle$ are shown. Similar to (a) the opposite-spin double occupation gets suppressed with $U/t$ while the equal-spin double occupation increases; an effect that is again enhanced by increasing the Hund's coupling $J$. 
		TPSC and DMFT obtain a qualitatively similar behaviour, with good agreement in the 
		same-orbital double occupation, which is suppressed by increasing interaction strength.
		For different orbitals the effect of the Hund's coupling of favoring a high-spin
		configuration becomes evident, but with increased deviation between TPSC and DMFT. We
		attribute this difference to the Hartree-Fock-like decoupling in the ansatz equations
		for the spin vertex (see eq.~\eqref{Eq: AnsatzGamma}) in TPSC.}
\end{figure*}

First, we present the double occupancy $\langle n_{\mu,\sigma}n_{\nu,\sigma'}\rangle$ in fig.~\ref{fig: nn} where TPSC results are compared to DMFT.

fig.~\ref{fig: nn}(a) shows that the double occupations $\langle n_{\mu,\sigma}n_{\mu,-\sigma}\rangle$ decrease when increasing the interaction values $U,J$  as expected from the local on-site interactions. 
This effect is further enhanced by enlarging the Hund's coupling $J$ which favors equal spin states in different orbitals. Moreover, in fig.~\ref{fig: nn}(b) and (c) the double occupations $\langle n_{\mu,\sigma}n_{\nu,\sigma'}\rangle$ drop by increasing $U/t$ when $\sigma\neq\sigma'$. This effect is again due to the on-site repulsion that penalizes double occupation. On the other hand, the equal spin double occupation $\langle n_{\mu,\sigma}n_{\nu,\sigma}\rangle$ increases with $U/t$ and is thus enhanced by increasing the Hund's coupling $J$. This is the counterpart to the decrease in $\langle n_{\mu,\sigma}n_{\nu,-\sigma}\rangle$. The results between both methods are comparable, except
at large $U/J$ (see fig.~\ref{fig: nn}(b) and (c)).
Specifically, while DMFT shows a drop in the equal spin double occupation $\langle n_{\mu,\sigma}n_{\nu,\sigma}\rangle$ at low $U$ and $J$ it is always enhanced compared to the noninteracting value 0.25 in TPSC. Since this drop in DMFT occurs at small values of the Hund's coupling
we conclude that the tendency to form a high-spin configuration is not yet dominating over the direct
intraorbital interaction $U' \approx U$ in the itinerant phase.
As $U$, and correspondingly $J$, is increased, the system
becomes more localized and a high spin state developes, as evident by the subsequent
increase in $\langle n_{\mu,\sigma}n_{\nu,\sigma}\rangle$. This effect is not 
captured in TPSC, which we attribute to the Hartree-Fock-like decoupling in the ansatz equations
for the spin vertex (see eq.~\eqref{Eq: AnsatzGamma}).

\begin{figure}
	\includegraphics[width=\columnwidth]{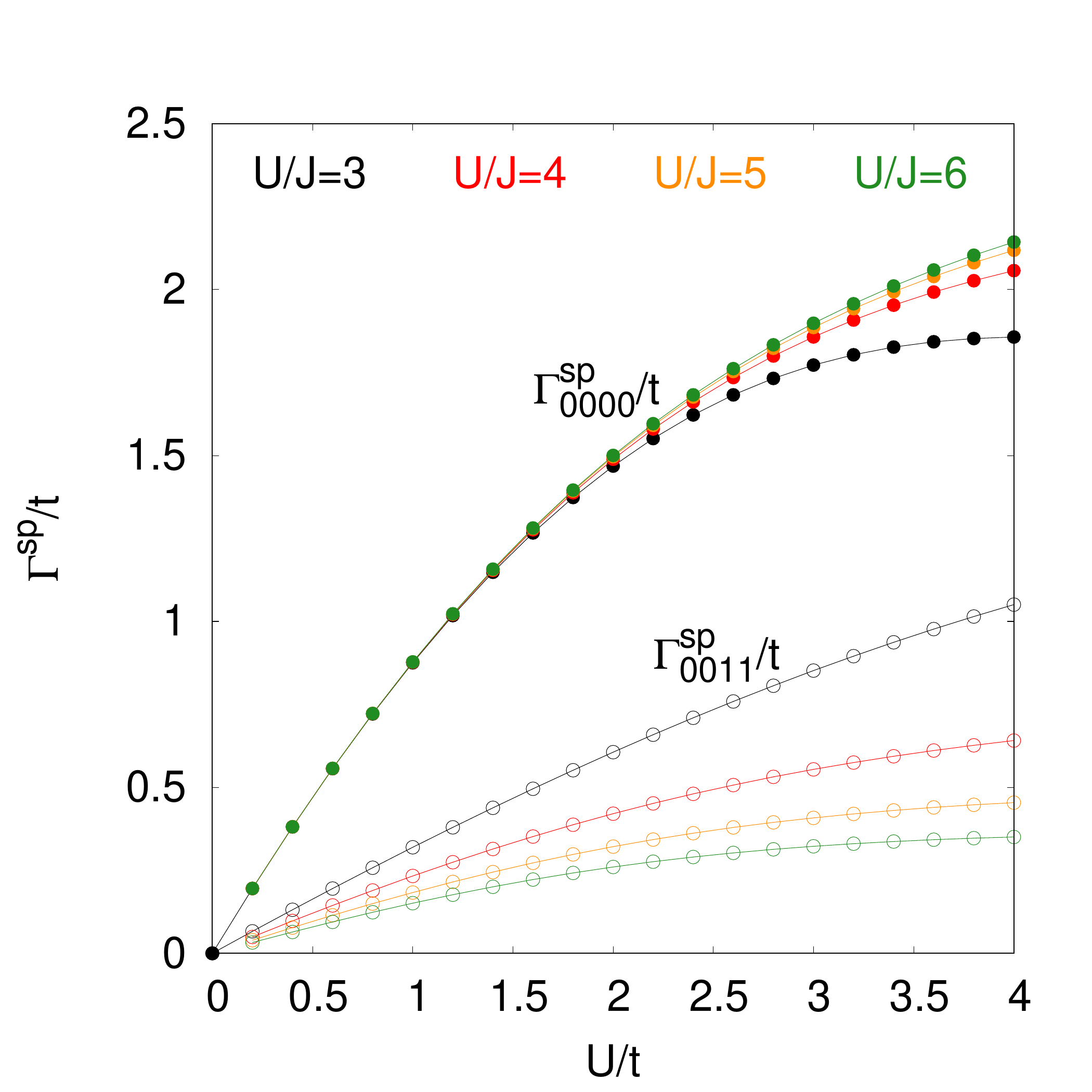}%
	\caption{\label{fig: Usp}
		The matrix elements of the spin vertex $\Gamma^{sp}$ are shown in dependence of $U/t$ where only non-zero matrix elements are shown. Moreover, we do not show $\Gamma^{sp}_{\mu\nu\mu\nu}=\Gamma^{sp}_{\mu\nu\nu\mu}$ because they are equal to $\Gamma^{sp}_{\mu\mu\nu\nu}$, $\mu\neq\nu$. As is in single-orbital TPSC, the (Kanamori-Brueckner) screening, i.e. the saturation of the spin vertex with increasing $U/t$, is observed. Further, it is evident that larger Hund's couplings suppress the matrix elements $\Gamma^{sp}_{\mu\mu\mu\mu}$ while the opposite is the case for $\Gamma^{sp}_{\mu\mu\nu\nu}$, $\mu\neq\nu$.}
\end{figure}

Next, we investigate the renormalized interaction matrices in the spin and charge channel $\Gamma^{sp}$ and $\Gamma^{ch}$. We start with $\Gamma^{sp}$ since it is determined first in the TPSC procedure. Note that under the constraint of intermediate interaction values $U/t$ and $J/t$ to ensure numerical stability, it is always possible to find $\Gamma^{sp}$ such that the local spin sum rules (eqs.~\eqref{Eq: spinsumrule1}, \eqref{Eq: spinsumrule2} and \eqref{Eq: spinsumrule3}) can be fulfilled exactly. The spin vertex $\Gamma^{sp}$ is shown in fig.~\ref{fig: Usp}.
Similar to single-orbital TPSC,  fig.~\ref{fig: Usp} shows the
Kanamori-Brueckner screening, namely the saturation of the spin vertex
$\Gamma^{sp}$ with increasing $U/t$. The screening of
$\Gamma^{sp}_{\mu\mu\mu\mu}$ is stronger (compare $U/J=3$ and $U/J=6$) if the
Hund's coupling is larger while the opposite is true for
$\Gamma^{sp}_{\mu\mu\nu\nu}$, $\mu\neq\nu$. This can be understood from the
fact that larger Hund's coupling favors double occupation of equal spins in 
different orbitals (see fig.~\ref{fig: nn}) while it suppresses double occupation of
opposite spins in the same orbital. This effect reduces the screening (see
eq.~\eqref{Eq: Gamma_local}) in $\Gamma^{sp}_{\mu\mu\nu\nu}$ if $\mu=\nu$ and
suppresses the screening if $\mu=\nu$.\\

\begin{figure*}
	\includegraphics[width=\textwidth]{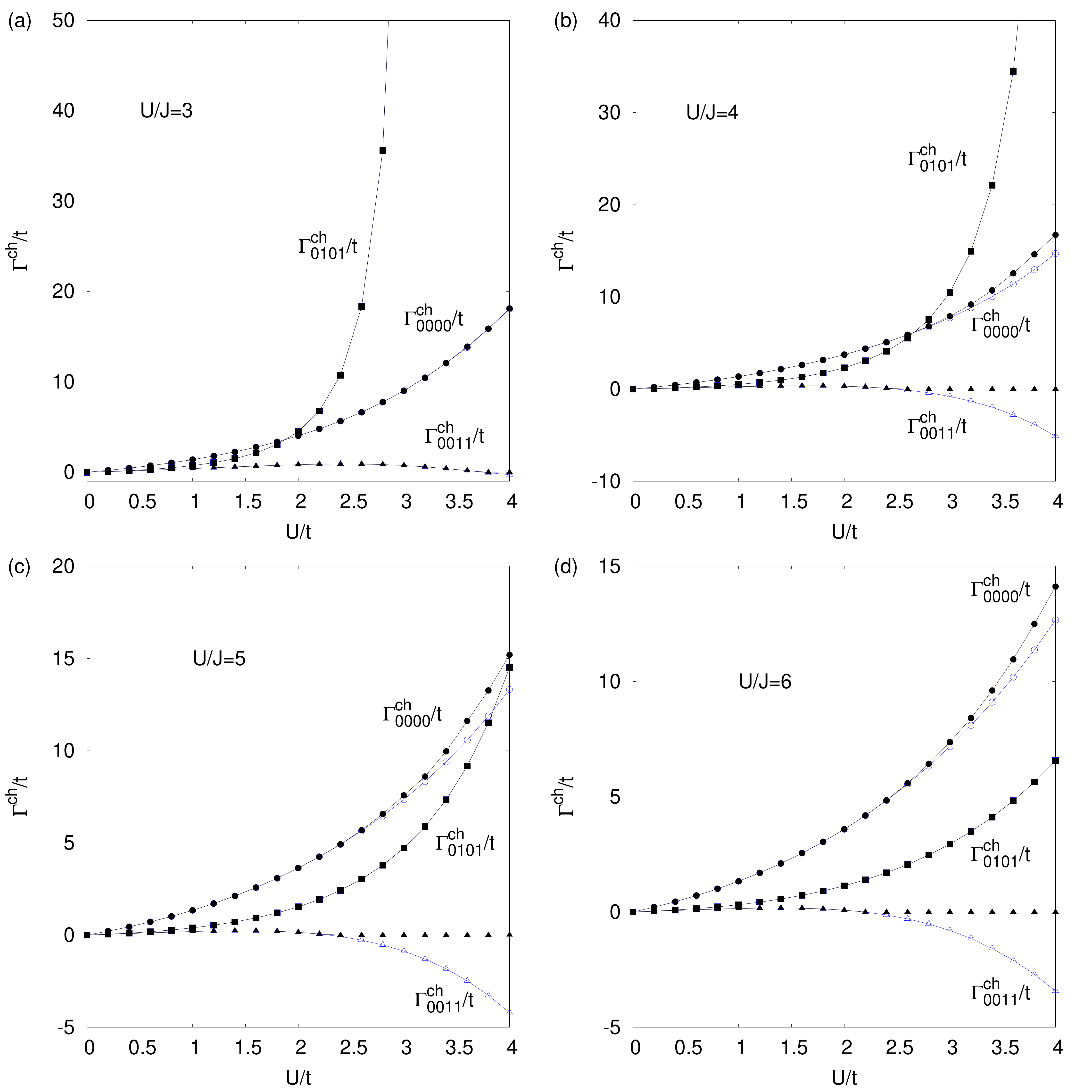}%
	\caption{\label{fig: Uch} 
		The matrix elements of the charge vertex $\Gamma^{ch}$ are shown in dependence of $U/t$. Redundant matrix elements that can be determined by the ones shown in the figure were left out. (a) to (d) show the same function for $U/J=3,4,5,6$ respectively. The open symbols that are mostly overlapped by their filled counterparts represent the solution for the charge vertex $\Gamma^{ch}$ without restriction while the filled symbols show the result when $\Gamma^{ch}$ is restricted to positive matrix elements. (a) and (b) show that for large Hund's coupling $J/U$ the matrix element $\Gamma^{ch}_{0101}$ diverges. On the other hand for small Hund's coupling the matrix element $\Gamma^{ch}_{0011}$ has the tendency to converge to negative solutions which contribute to a negative spectral function $A(\vec k,\omega)$ that is unphysical. Therefore, we restrict $\Gamma^{ch}$ to positive matrix elements and minimize the difference between right-hand side and left-hand side of the local charge sum rule equations (eqs.~\ref{Eq: charge_sumrules}). The results for this constrained calculations are the filled symbols. The general tendency of the charge vertex is similar to the single-orbital TPSC with a diverging behavior which leads to the suppression of charge fluctuations in the system.}
\end{figure*}

\begin{figure}
	\includegraphics[width=\columnwidth]{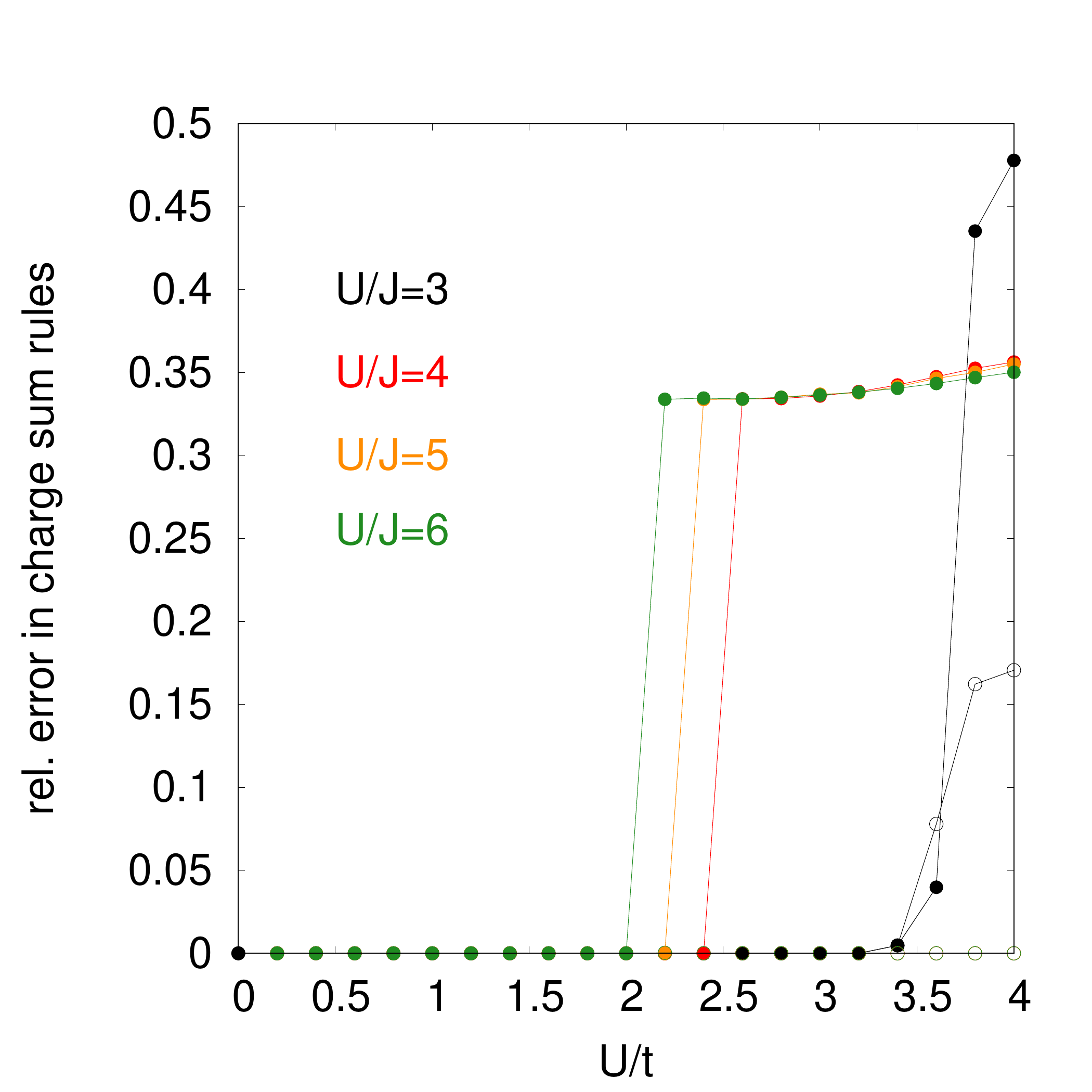}%
	\caption{\label{fig: ch_err}
		The relative error in the charge sum rules for a constrained determination of $\Gamma^{ch}$ is shown in dependence of $U/t$. In the case of small interaction values $U/t$, $\Gamma^{ch}$ can be determined such that the local charge sum rules are fulfilled exactly. Only in the cases of large Hubbard interaction $U/t$ in combination with small Hund's coupling to Hubbard interaction ratio $J/U$ we observe deviations up to 50\% or in the case where $\Gamma^{ch}_{\mu\mu\nu\nu}$ undergoes a sign change (see fig.~\ref{fig: Uch}) a clear jump in the error happens at $U/t=2.6$.}
\end{figure}

Having obtained the spin vertex $\Gamma^{sp}$ and the double occupations $\langle n_{\mu,\sigma}n_{\nu,\sigma'}\rangle$ the next step in the TPSC procedure is to determine the charge vertex from the local charge sum rules~(eqs.~\eqref{Eq: charge_sumrules}). 
The calculation of $\Gamma^{ch}$ from the local charge sum rules is doable in a systematic way but, unfortunately, is not straightforward and this might have several reasons as we will elucidate in the following discussion.\\
We first consider an unconstrained numerical solution of the local charge sum rules and show the obtained results as open symbols -- those are mostly overlapping the same filled symbol -- in fig.~\ref{fig: Uch}.
The unconstrained calculations have two important issues in the cases of large Hund's coupling $J$ and in the case of large Hubbard interactions $U$. In the first case we observe that $\Gamma^{ch}_{\mu\nu\mu\nu},~\mu\neq \nu,$ has the tendency of strong divergence which could possibly make the TPSC procedure numerically unstable due to overflow or precision errors when handling diverging matrix elements. 
In this study the diverging charge vertex components went up to values $~\sim 10^{11}$ and we had no numerical problems but we cannot exclude the previously mentioned errors in the general case.
On the other hand fig.~\ref{fig: Uch} (b)-(d) shows that for lower $J/U$ values the matrix element $\Gamma^{ch}_{\mu\mu\nu\nu},~\nu\neq\mu,$ has the tendency to go to negative values. This gives a negative contribution to the spectral function $A(\vec k,\omega)$ and must be avoided.\\
To do this, one imposes numerically the constraint of positive solutions and computes $\Gamma^{ch}$ such that the difference between left-hand and right-hand side of equations~\eqref{Eq: charge_sumrules} is minimal.\\
The result of this restricted charge vertex calculation is shown as filled symbols in fig.~\ref{fig: Uch}. As one can see this has in general only an effect if $U/t$ is large and $J/U$ is small. In the other cases the unconstrained determination of $\Gamma^{ch}$ gives purely non-negative results -- except for the point at $U/t=3.8,~4.0$ in fig.~\ref{fig: Uch}(a) which is due to the divergence of $\Gamma^{ch}_{0101}$. Moreover, fig.~\ref{fig: Uch} shows that the effect of the restriction has a small impact on the matrix elements $\Gamma^{ch}_{0101}$ and $\Gamma^{ch}_{0000}$. As in \onlinecite{Miyahara2013} we also 
find that in the cases of small Hund's coupling $J$ compared to $U$ the matrix element $\Gamma^{ch}_{0011}$ is small compared to the other matrix elements and this further corroborates the procedure to set those matrix elements to zero.\\
Finally, we remark the general tendency of the charge vertex to diverge as a function of increasing $U/t$. 
This leads -- as in the single-orbital case -- to the freezing of charge fluctuations in the system.\\
This non-perturbative divergence in the charge channel was originally interpreted as a precursor effect of the Mott transition~\onlinecite{VilkTremblay1997}, that is not only observed in TPSC but also in DMFT~\onlinecite{Schaefer2013} and DCA~\onlinecite{Gunnarsson2016}.
More recent studies~\onlinecite{Chalupa2018} were able to show that such divergences can also occur independent of the Mott transition and that the formation of local magnetic moments can also lead to such divergences~\onlinecite{Chalupa2020}.\\
Indeed, the divergence of the charge vertex can lead to interesting physical consequences such as phase-separation instabilities as observed in DMFT~\onlinecite{Nourafkan2019,Reitner2020}.\\
Since those charge vertex divergences are an interesting and active field of research we give a sketch of how they appear within TPSC.
For simplicity we go back to the single-band TPSC where one can
see from the Bethe-Salpeter equation (eq.~\eqref{eq: chi_sp_1band}), the local spin sum rule (eq.~\eqref{eq: sp_sumrule_1band}) and the ansatz eq.~\eqref{eq: Usp_ansatz_1band} that the double occupation $\langle n^{}_{\uparrow}n^{}_{\downarrow}\rangle$ is determined as the root of the equation
\begin{equation}
\frac{1}{\tilde\beta N_{\vec q}}\sum_{q} \frac{2\chi^{0}(q)}{1-U\frac{\langle n^{}_{\uparrow}n^{}_{\downarrow}\rangle}{\langle n^{}_{\uparrow}\rangle\langle n^{}_{\downarrow}\rangle}\chi^{0}(q)} = 2\langle n^{}_{\uparrow}\rangle - 2\langle n^{}_{\uparrow}n^{}_{\downarrow}\rangle.
\end{equation}
Obviously, an increasing value of Hubbard interaction $U$ will lead to an increase on the left-hand side of the local spin sum rule hence the double occupation $\langle n^{}_{\uparrow}n^{}_{\downarrow}\rangle$ has to decrease in order
to readjust equality because the non-interacting susceptibility is positive and real.
If we consider the decreasing double occupation in the local charge sum rule,
\begin{equation}
\frac{1}{\tilde\beta N_{\vec q}}\sum_{q} \frac{2\chi^{0}(q)}{1+\chi^{0}(q)\Gamma^{ch}} = \langle n^{}\rangle + 2\langle n^{}_{\uparrow}n^{}_{\downarrow}\rangle - \langle n\rangle^2,
\end{equation}
we easily conclude that the charge vertex has to increase to compensate for the
loss on the right-hand side.\\ 
In the case of half-filling and the limit where
the double occupation goes to zero,  the right-hand side of the local
charge sum rules also vanishes and thus necessitates a divergence of the charge
vertex.\\
Within TPSC the feature of the diverging charge vertex is therefore intimately
connected with the Pauli principle from which the local sum rules were
derived.
\\~\\
Since the restriction to positive matrix elements of $\Gamma^{ch}$ will necessarily lead to an error in the local charge sum rule it is worth to take a look at this error. The relative error of the local charge sum rule is defined as the sum of differences between right-hand side and left-hand side of the equations~\eqref{Eq: charge_sumrules} divided by the respective right-hand sides. Those errors are shown in fig.~\ref{fig: ch_err}.
Clearly, the plot shows that for small interactions $U/t$ the constrained and unconstrained determination yield the same (non-negative) $\Gamma^{ch}$ that fulfills the local charge sum rules exactly. 
On the one hand, if the Hubbard interaction is large compared to the hopping amplitude $t$ ($U/t\geq 3.5$) and to the Hund's coupling $J$ ($U/J\geq 3$), large deviations appear that rise close to $50\%$. The largest deviations, as can be already guessed from fig.~\ref{fig: Uch}, happen where the restriction induces largest differences, namely $\Gamma^{ch}_{\mu\mu\nu\nu},~\mu\neq\nu$ and therefore the largest deviations appear in the $\mu\mu\nu\nu$-charge sum rule. 
This also explains the jump in the error at $U/t=2.6$ where $\Gamma^{ch}_{\mu\mu\nu\nu}$ changes sign and the local charge sum rules cannot be fulfilled anymore. 
The fact that even the unconstrained solutions for $\Gamma^{ch}$ can lead to an error of up to $15\%$ as for $U/J=3$ at $U/t=3.8,~4$ is due to the numerical instability of the diverging charge vertex.\\
Although these numbers seem rather discomforting one has to keep in mind that they only occur at large interaction values $U/t$ ($U/t\geq 3.5$) and in those regions it is the spin susceptibility that dominates the contribution in the self-energy equation~\eqref{eq: Sigma_final}. One can therefore expect to obtain a self-energy that is still qualitatively and even quantitatively accurate as long as the original assumption of this method, namely that $\Gamma^{sp/ch}$ are constant, is still a good approximation. 

\begin{figure}
	\includegraphics[width=\columnwidth]{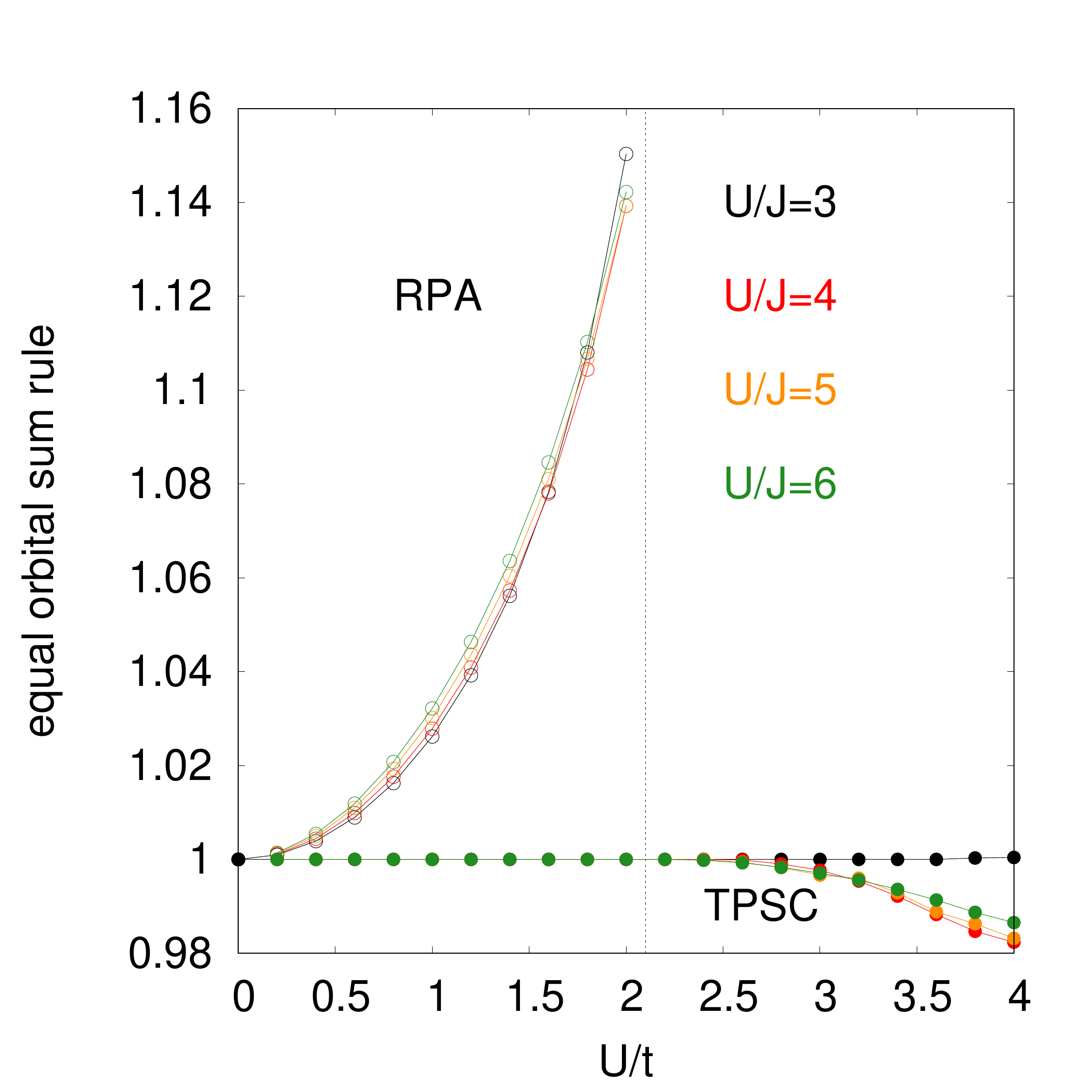}%
	\caption{\label{fig: err_TPSC_RPA} 
		The value of the equal orbital sum rule (eq.~\eqref{Eq: equal_orbital_sum_rule}) as a function of $U/t$ for different values of $U/J$ is presented for RPA (open symbols) and TPSC (filled symbols). TPSC by construction fulfills the equal orbital sum rule except for the case of large $U/t$ where the restriction of $\Gamma^{ch}$ leads to small deviations. On the other hand, RPA starts to violate the sum rule already at small interaction values $U/t$ until it reaches the magnetic instability at around $U/t~\approx 2.2$.}
\end{figure}

To conclude the discussion of the spin and charge vertices $\Gamma^{sp/ch}$, it is worthwhile to investigate the degree of local spin and charge sum rules' violation if one does not renormalize the spin and charge vertex, i.e. taking the RPA values from eqs.~\eqref{Eq: Gamma_sp_RPA} and \eqref{Eq: Gamma_ch_RPA}. Since the sum rules (eqs.~\eqref{Eq: charge_sumrules}, \eqref{Eq: spinsumrule1}, \ref{Eq: spinsumrule2} and \ref{Eq: spinsumrule3}) all depend on double occupations that are in principle another unknown, we look at the sum of charge and spin sum rules in the case where all indices are equal, i.e.
\begin{equation}
\frac{1}{\tilde\beta N_{\vec q}}\sum\limits_{\vec q,iq_n}\left(\chi^{sp}+\chi^{ch}\right)_{\mu\mu\mu\mu}(\vec q,iq_n)= 2\langle n_\mu\rangle-\langle n_\mu\rangle ^2 = 1,\label{Eq: equal_orbital_sum_rule}
\end{equation}
where we have used that our calculations are at half filling, $\langle n_\mu\rangle=1$.\\
The result of the comparison is shown in fig.~\ref{fig: err_TPSC_RPA}.
TPSC fulfills the equal orbital sum rule by construction up to the largest values of $U/t$ considered, namely $U/t=4$. At $U/t>2.6$ a small deviation becomes visible which is due to the fact that in this regime one has to restrict $\Gamma^{ch}$ which leads to small deviations in all local charge sum rules and consequently also in the equal orbital sum rule (eq.~\ref{Eq: equal_orbital_sum_rule}). 
On the other hand, this same calculation once more shows that the largest deviations due to the constrained determination of $\Gamma^{ch}$ happen in the $\mu\mu\nu\nu$- sum rule where $\mu\neq\nu$ which is not part of the equal orbital sum rule and therefore the error is still very small compared to the ones presented in fig.~\ref{fig: ch_err}.\\
Nonetheless, fig.~\ref{fig: err_TPSC_RPA} demonstrates that no renormalization of the spin and charge vertex (RPA) can lead to strong violation of the equal orbital sum rule and therefore to a specific aspect of the Pauli principle. It was shown in Ref.~\onlinecite{VilkTremblay1997} that this deviation is quadratic when the interaction parameters are small and we reproduce the same result in the multi-orbital case.

\begin{figure}
	\includegraphics[width=\columnwidth]{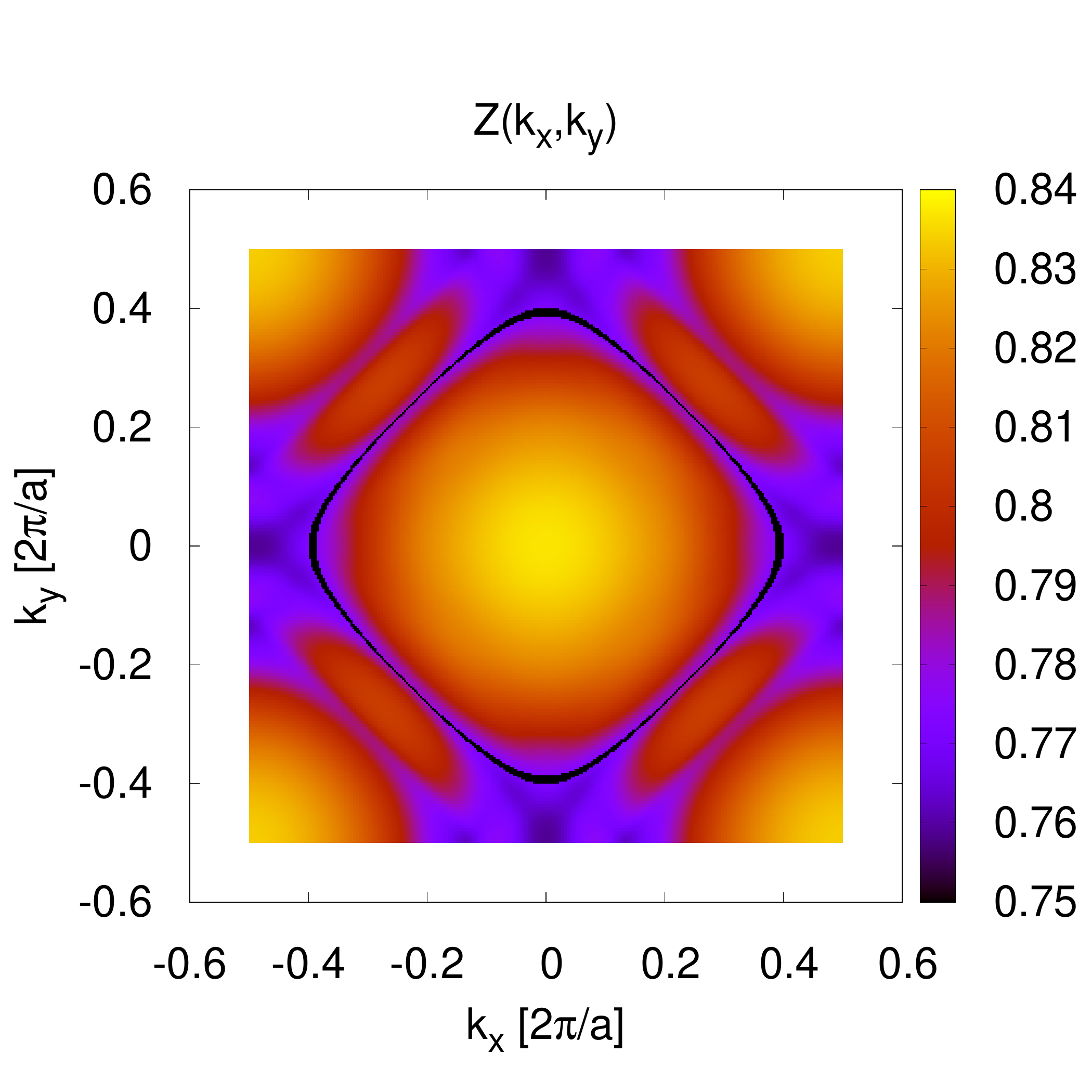}%
	\caption{\label{fig: QP_weight}
		Quasi-particle spectral weight $Z(k_x,k_y)$ for $U/t=2.5$ and $U/J=4$, $\langle n\rangle=0.8$ and $k_BT/t = 0.03$. The quasi-particle weight shows a strong momentum dependence where the minima are located along in the vicinity of the four hot spots along the lines $k_x=0$ and $k_y=0$ lines. $Z$ was obtained by linear extrapolation of the imaginary part of $\Sigma(k,i\omega_n)$ to $i\omega_n=0$. In black we show the Fermi surface of the system.}
\end{figure}

\begin{figure}
	\includegraphics[width=\columnwidth]{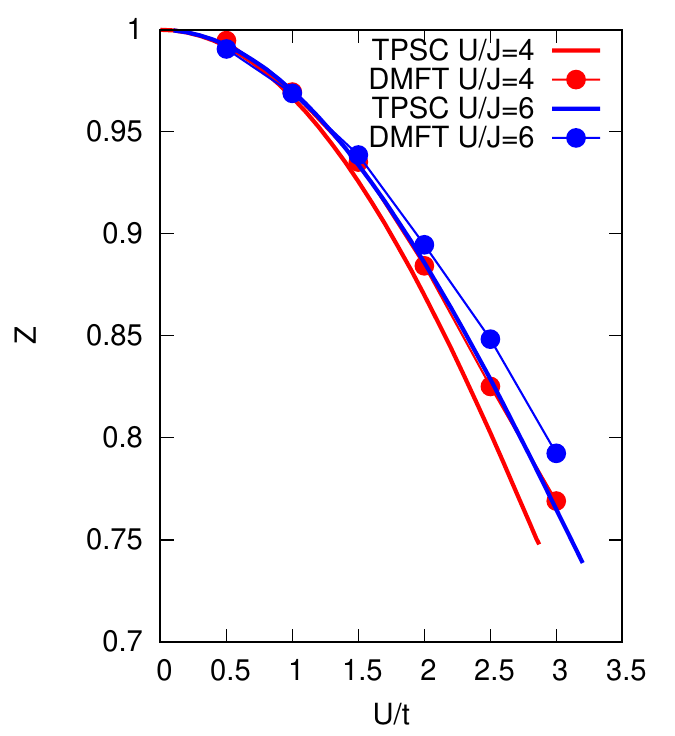}%
	\caption{\label{fig: Z_loc} 
		Local quasi-particle spectral weight $Z$  is obtained from the local limit of the TPSC self-energy $\Sigma(i\omega_n)$ (see main text for definition). TPSC and DMFT agree well in the parameter range considered.}
\end{figure}

\begin{figure}
	\includegraphics[width=\columnwidth]{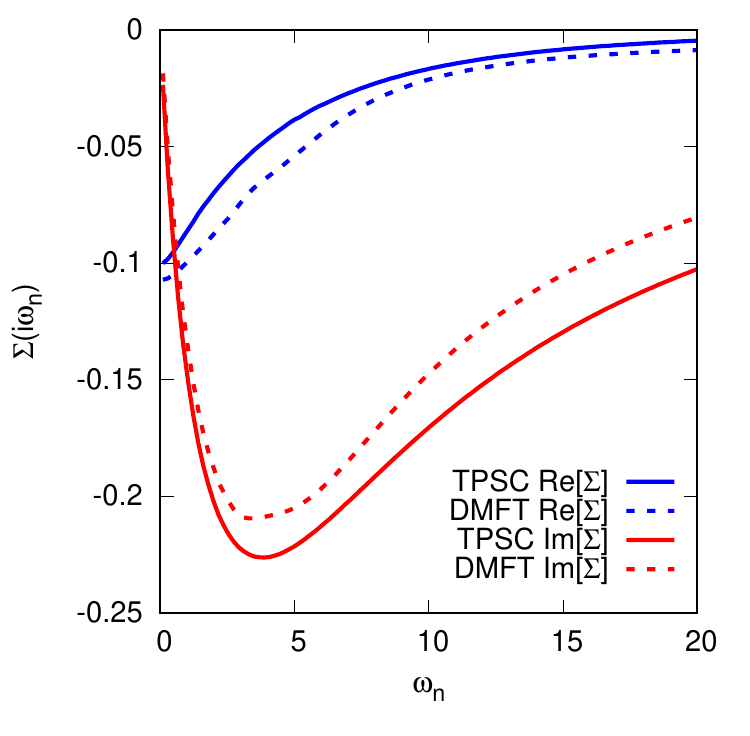}%
	\caption{\label{fig: Sigma_iwn} 
		Real and imaginary part of the local limit of the self-energy $\Sigma(i\omega_n)$ from TPSC and DMFT. Except for the large $\omega_n$-limit of the imaginary part of the self-energy TPSC and DMFT agree well. This discrepancy is due to the missing frequency dependencies of the spin and charge vertices $\Gamma^{sp}$ and $\Gamma^{ch}$.}
\end{figure}

Next, we consider the results of the self-energy where we have picked $k_BT/t=0.03$ and $\langle n\rangle =0.8$ for all calculations.
Note that we dropped the orbital index of all objects in the following since the self-energy and the Green's function are diagonal and orbital degenerate in orbital-index space.
This means that all presented functions are always computed from orbital averages of the diagonal matrix elements.\\
For all the following results the deviation from left-hand to right-hand side of the internal accuracy check equation~\eqref{Eq-tr-Sigma-G} was always lower than 2.5\% for both $G$ and $G_0$.\\
For $U/t=2.5$ and $U/J=4$ the spin fluctuations in the system are already large and the momentum-dependence of the self-energy leads to strongest suppression of quasi-particle spectral weight $Z(\vec k)$ at the hot spots where the number drops down to 0.75 (see fig.~\ref{fig: QP_weight}).
To compare the momentum dependent quasi-particle spectral weight $Z(\vec k)$ to the momentum independent result of DMFT we go into the local limit of TPSC by replacing $\Sigma(\vec k,i\omega_n)$ by its momentum average $\Sigma(i\omega_n):=\frac{1}{N_{\vec k}}\sum_{\vec k}\Sigma(\vec k,i\omega_n)$. The result of this comparison of quasi-particle weights of TPSC and DMFT is shown in fig.~\ref{fig: Z_loc}. 
Moreover, we compared the self-energy for $U/t=2.5$ and $U/J=4$ of DMFT and the local limit of TPSC in fig.~\ref{fig: Sigma_iwn}.
We observe that the momentum averaged TPSC and DMFT are in good agreement and show the same qualitative trend. The largest discrepancy is in the imaginary part of the self-energy at large Matsubara frequencies $\omega_n$ which is due to the fact that the renormalized spin and charge vertices $\Gamma^{sp},~\Gamma^{ch}$ do not rescale to the bare values -- since the vertices are constant -- to give the right constant in the $1/i\omega_n$ expansion of the self-energy.  

\section{Summary and Outlook}
\label{sec: Outlook}
We reviewed the multi-orbital extensions of the Two-Particle Self-Consistent approach that was originally 
formulated by Vilk and Tremblay~\onlinecite{VilkTremblay1997} for the single-orbital case. This approach is able to enforce physical properties of the system such as the Pauli principle, the Kanamori-Brueckner screening, the Mermin-Wagner theorem and many conservation laws that are inherent to conserving approximations. 
However, deviations can occur by increasing the interaction strength $U/t$ since by construction one considers the spin and charge vertices $\Gamma^{sp},~\Gamma^{ch}$ to be orbital-dependent constants and this is not compatible with all local charge and spin sum rules (see eqs.~\eqref{Eq: charge_sumrules}),
\eqref{Eq: spinsumrule1}, \eqref{Eq: spinsumrule2}, and  \eqref{Eq: spinsumrule3}).\\
Although this constraint leads to deviations of up to $50\%$ in the local charge sum rules, the latter has usually a small effect on the self-energy since the spin fluctuations are the major contributions in repulsive Hubbard models. 
All self-energy calculations in this review showed a deviation in the internal accuracy equation~\eqref{Eq-tr-Sigma-G} of at most $1.5\%$. 
Most importantly for the self-energy calculations we saw that indeed the quasi-particle weight $Z(\vec k)$ shows momentum dependence where the strongest suppression of spectral weight occurs close to the $X$ and $Y$ point.\\
TPSC is therefore a close relative of conserving approximations that not only respects the local spin and charge sum rules and the $\text{tr}(\Sigma G)$ sum rule to high degree but does also not violate the Mermin-Wagner theorem. Moreover, due to the convolution expressions (eqs.~\eqref{Eq: chi0} and \eqref{eq: Sigma_final}) it is very fast to perform TPSC calculations via Fast Fourier Transform and the convolution theorem. In addition, new numerical techniques were developed to further improve numerical efficiency~\onlinecite{Shinaoka2017,wang2020,Li2020}.
The method provides momentum- and frequency-dependent self-energies in the regime of weak to intermediate coupling strength and was successfully applied to multi-band high-Tc superconductors~\onlinecite{Miyahara2013,Zantout2019,Bhattacharyya2020}.  

As for future work it might be interesting to extend multi-orbital TPSC to study further
neighbor interactions as was already done for the single-orbital
TPSC~\onlinecite{Davoudi2006,Davoudi2007,Davoudi2008} or superconductivity as was
already started in~\onlinecite{Miyahara2013}. From the method point of view it would
be also worthwhile to incorporate at least some kind of frequency dependence to
the spin and charge vertices $\Gamma^{sp},~\Gamma^{ch}$ to be able to fulfill
the local spin and charge sum rules to a higher degree and get the right high
frequency behavior of the imaginary part of the self-energy. This would also
improve the method in the sense that Hubbard physics would be visible.
This is especially desirable because the feature of diverging charge vertices has
attracted a lot of attention in recent years and contains itself interesting physics~\onlinecite{Schaefer2013,Gunnarsson2016,Chalupa2018,Nourafkan2019,Chalupa2020,Reitner2020}.\\
Moreover, one could include self-energy contributions from the transversal channel like in~\onlinecite{Moukouri2000,Allen2001} which was found numerically to improve on the tail of the self-energy. 
Unfortunately, the form of $\Gamma^{sp,0}$ does not have the simple form of equation~\eqref{Eq: Gamma_sp0} and one cannot straightforwardly set $\Gamma^{sp}_{\mu\mu\nu\nu}=\Gamma^{sp}_{\mu\nu\mu\nu}=\Gamma^{sp}_{\mu\nu\nu\mu}$ as we did for the longitudinal channel. This also means that there are not enough local spin sum rules and ansatz equations to determine $\Gamma^{sp}$.

Alternatively, it may be worthwhile to use double occupations from precise methods and calculate $\Gamma^{sp}$ from the local spin sum rules without need of an ansatz equation. This procedure was already suggested and used in the single-orbital case in~\onlinecite{Kyung2003}.

\textbf{Acknowledgments}
The authors thank Ryotaro Arita, Chlo{\'e}-Aminata Gauvin-Ndiaye, Peter Hirschfeld and Andr{\'e}-Marie Tremblay for fruitful discussions. R.V. acknowledges the Deutsche Forschungsgemeinschaft (DFG, German Research Foundation) for funding through TRR 288 - 422213477 (project B05).

\clearpage

\bibliography{literature_arxiv}
\bibliographystyle{apsrev4-1}
\end{document}